\documentclass[fleqn,usenatbib]{mnras}

\usepackage{newtxtext,newtxmath}

\usepackage[T1]{fontenc}
\usepackage{ae,aecompl}

\def\be{\begin{equation}}
\def\ee{\end{equation}}
\def\CF3{{\sc cosmicflows-3}}
\usepackage{rotating}
\usepackage{multirow}
\usepackage{graphicx}	
\usepackage{amsmath}	
\usepackage{amssymb}	

\title[Cosmic flow in 2MTF and CF3]{Bulk flow and shear in the local Universe: 2MTF and \CF3}

\author[F. Qin et al.]{
Fei Qin$^{1,2}$\thanks{E-mail: fei.qin@research.uwa.edu.au},
Cullan Howlett$^{1,2}$,
Lister Staveley-Smith$^{1,3}$,
Tao Hong$^{2,4,5}$
\\
$^{1}$International Centre for Radio Astronomy Research (ICRAR), University of Western Australia, 35 Stirling Hwy, Crawley, WA 6009, Australia\\
$^{2}$ARC Centre of Excellence for All-sky Astrophysics (CAASTRO)\\
$^{3}$ARC Centre of Excellence for All Sky Astrophysics in 3 Dimensions (ASTRO 3D)\\
$^{4}$National Astronomical Observatories, Chinese Academy of Sciences, 20A Datun Road, Chaoyang District, Beijing 100101, China\\
$^{5}$CAS Key Laboratory of FAST, National Astronomical Observatories, Chinese Academy of Sciences
}

\date{Accepted XXX. Received YYY; in original form ZZZ}

\pubyear{2015}

\begin{document}
\label{firstpage}
\pagerange{\pageref{firstpage}--\pageref{lastpage}}
\maketitle

\begin{abstract}
  
The low-order kinematic moments of galaxies, namely bulk flow and shear, enables us to test whether theoretical models can accurately describe the evolution of the mass density field in the nearby Universe. 
We use the so-called $\eta$MLE maximum likelihood estimator in log-distance space to measure these moments 
from a combined sample of the 2MASS Tully-Fisher (2MTF) survey and the \CF3 (CF3) compilation. Galaxies common between 2MTF and CF3 demonstrate a small zero-point difference of $-0.016\pm 0.002$ dex.  
We test the $\eta$MLE on 16 mock 2MTF survey catalogues in order to explore how well the $\eta$MLE recovers the true moments, and the effect of sample anisotropy.
On the scale size of 37 $h^{-1}$ Mpc, we find that the bulk flow of the local Universe is $259\pm 15$ km s$^{-1}$ in the direction is $(l,b)=(300\pm 4^{\circ},  23\pm 3^{\circ})$ (Galactic coordinates). The average shear amplitude is $1.7\pm 0.4$ $h$ km s$^{-1}$ Mpc$^{-1}$. We use a variable window function to explore the bulk and shear moments as a function of depth. In all cases, the measurements are consistent with the predictions of the $\Lambda$ cold dark matter ($\Lambda$CDM) model.

\end{abstract} 




\section{Introduction}

In the local Universe, the gravitational effects of mass density fluctuations exert perturbations on galaxies' redshifts on top of Hubble's Law, called `peculiar velocities'. The dipole and the quadruple of the peculiar velocity field, namely `bulk flow' and `shear' respectively, enable us to trace the matter density fluctuations and test whether   
the cosmological model accurately describes the motion of galaxies in the nearby Universe.

In previous work related to the measurement of the bulk and shear moments 
\citep{1989MNRAS.241..787S,1995ApJ...455...26J,1998ApJ...507...64W,2001AstL...27..765P,2010MNRAS.407.2328F,2014MNRAS.445..402H,2016MNRAS.455..386S,2018MNRAS.477.5150Q}, the results largely agree with the $\Lambda$CDM prediction. However some studies have measured large values for the bulk flow, in apparent disagreement  with the $\Lambda$CDM prediction. For example, \citet{2009MNRAS.392..743W} measure 407 $\pm$ 81 km s$^{-1}$ on the scale size of 50 $h^{-1}$ Mpc.

The bulk and shear moments
are usually measured in velocity space ($v$-space) or log-distance space ($\eta$-space). In $v$-space, the main measurement techniques are \citep{1988MNRAS.231..149K,2007MNRAS.375..691S,2009MNRAS.392..743W,2010MNRAS.407.2328F,2014MNRAS.445..402H}: log-linear
$\chi^2$ minimization, minimum variance (MV) estimation and maximum likelihood estimation (MLE).
Some of these $v$-space estimation techniques
assume that the measured peculiar velocities have Gaussian errors, which is not the case for the usual estimator of peculiar velocity. \cite{2015MNRAS.450.1868W} therefore introduced a peculiar velocity estimator which has Gaussian errors and, under some circumstances is unbiased. Alternatively, as shown by previous authors including \citet{1995MNRAS.276.1391N,2011ApJ...736...93N} and \citet{2018MNRAS.477.5150Q}, the bulk and shear moments in the local Universe can be measured in $\eta$-space using the `$\eta$MLE' technique.
\cite{2011ApJ...736...93N} convert the model bulk flow into magnitudes analytically, using linear approximations, then convert to log-distance ratio and compare to the measurements, while \cite{2018MNRAS.477.5150Q}, convert the model bulk flow into log-distance ratio numerically without any approximations, then compare to the measurements.

In this work, we extend the $\eta$MLE in \cite{2018MNRAS.477.5150Q} to quadrupole (shear) measurements and, through weighting functions, compare the measured shear moments with $\Lambda$CDM prediction at different depths. We measure the bulk and shear moments from the combined dataset of \CF3 (CF3; \citealt{2016AJ....152...50T}) and 2MTF \citep{2014MNRAS.445..402H}.

The paper is structured as follows: in Section \ref{sec:data}, we introduce the data: 2MTF, CF3 and their combination. The theory associated with the low-order moments (bulk and shear) is introduced in Section \ref{sec:BKQtheory}. In Section \ref{sec:theory}, we summarize how these are estimated from the data. In Section \ref{sec:mocktests}, we discuss the bulk and shear moments obtained from the 2MTF mocks. The final results are presented in Section \ref{sec:dis}. We provide a conclusion in Section \ref{conc}.

This paper assumes spatially flat cosmology with parameters from the \cite{2014A&A...571A..16P}: $\Omega_m=0.3175$, $\sigma_8=0.8344$, $\Omega_{\Lambda}=0.6825$ and $H_{0} = 100$ $h$ km s$^{-1}$ Mpc$^{-1}$. We use these parameters to calculate the expected $\Lambda$CDM bulk flow and shear as well as the comoving distances.

\section{DATA} \label{sec:data}

 \subsection{CF3 and 2MTF}

\CF3 (CF3) is a full-sky compilation of distances and velocities \citep{2016AJ....152...50T}, containing 17\,669 galaxies reach $cz = 34\,755$ km~s$^{-1}$.
The data sources are heterogeneous, and include distances obtained from the luminosity-linewidth (Tully-Fisher) relation, the Fundamental Plane (FP), surface-brightness fluctuations, from Type Ia supernova (SNIa) observations, the tip of the Red Giant Branch (TRGB), with the largest recent increment being the FP sample of the Six-degree-Field Galaxy Survey (6dFGS) of \citet{2014MNRAS.445.2677S}.
We removed those galaxies with CMB frame redshift lower than 600 km s$^{-1}$, leaving 17\,407.

 2MTF is a Tully-Fisher
sample derived from 
the Two Micron All-Sky Survey. 
The Tully-Fisher relation is measured using 
H {\small I} rotation widths \citep{2005ApJS..160..149S,2011AJ....142..170H,2013MNRAS.432.1178H,2014MNRAS.443.1044M} for galaxies at redshifts measured in the 2MASS Redshift Survey \citep{2012ApJS..199...26H}.
The final 2MTF catalogue contains 2\,062 galaxies with a redshift cut 600 km~s$^{-1}<cz< 1.2 \times10^{4}$ km~s$^{-1}$. The 2MTF $K$-band magnitude limit is $11.25$ mag.

 \subsection{The combination of CF3 and 2MTF}

The combination of CF3 and 2MTF data offers the following advantages. Firstly, the combined data set is much deeper than 2MTF alone (CF3 extends out to three times the redshift of 2MTF). Secondly, the combined data set is more isotropic than CF3 alone (the projected sky density of CF3 is greater in the southern sky by a factor of 2.4, and the projected density of 2MTF is greater in the northern sky by 1.6).

 
In order to find the common galaxies in the two catalogue and calibrate out any zero-points, we need to cross-compare the estimated distances in the 2MTF and the CF3 data.
The `logarithmic distance ratio' for a galaxy, $\eta$ is defined as
\be\label{logd}
\eta \equiv \log_{10}\frac{d_z}{d_h}~,
\ee
where $d_z$ is the apparent distance of a galaxy, and is inferred from the observed redshift of the galaxy. The true comoving distance, $d_h$ is calculated from a redshift-independent measurement of the galaxy \citep{1995PhR...261..271S}. 2MTF uses the Tully-Fisher distance estimator, while CF3 uses a compilation of Fundamental Plane, Tully-Fisher and Type Ia supernovae. The CF3 catalogue does not have log-distance ratio data, but it lists distance modulus $\mu$, corresponding to $h=0.75$. We convert $\mu$ to $\eta$, assuming $h=0.75$, and assign an error for $\eta$ corresponding to $1/5$ of the error for $\mu$.

\begin{figure}  
 \includegraphics[width=\columnwidth]{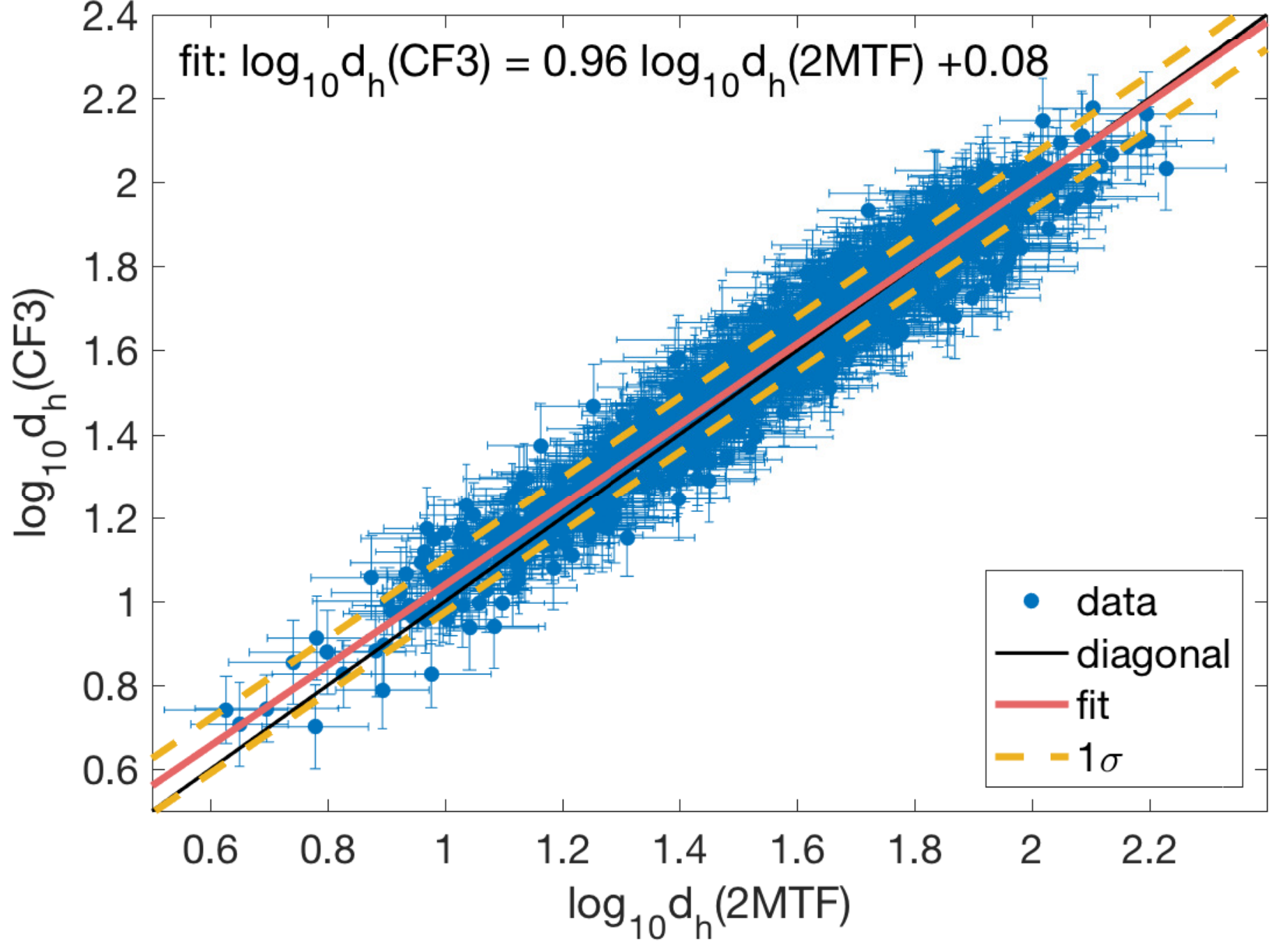}
 \caption{Comparing the CF3 distances to the 2MTF distances for 1\,096 common galaxies. The expected 1:1 relation for perfect agreement is shown in the solid black line. The {\tiny  HYPERFIT} line is shown in the solid red line. The $\pm1\sigma$ is indicated by the yellow dashed lines, and $\sigma=0.07$. }
\label{commongals}
\end{figure}

There are 1\,117 common galaxies in the 2MTF and the CF3. 
These galaxies are identified as having CF3 and 2MTF heliocentric velocity differences $|\Delta v_{hel}| < 150$ km s$^{-1}$. The CF3 distance estimator for these galaxies is mostly Tully-Fisher.
For each galaxy, we calculate $\log_{10}d_h$(2MTF) and $\log_{10}d_h$(CF3) then apply a linear fit with a $3\sigma$ clip. This removes 21 galaxies, leaving 1\,096 (we used the {\tiny  HYPERFIT} package; \citealt{2015PASA...32...33R}). In Fig.\ref{commongals}, we plot $\log_{10}d_h$(2MTF) against $\log_{10}d_h$(CF3) for these galaxies. The average difference is
\be
\Big\langle\log_{10}\frac{d_h(2MTF)}{d_h(CF3)}\Big\rangle=-0.016\pm0.002~,
\ee
representing a 4 per cent difference in distance.

\begin{figure*} 
 \includegraphics[width=175mm]{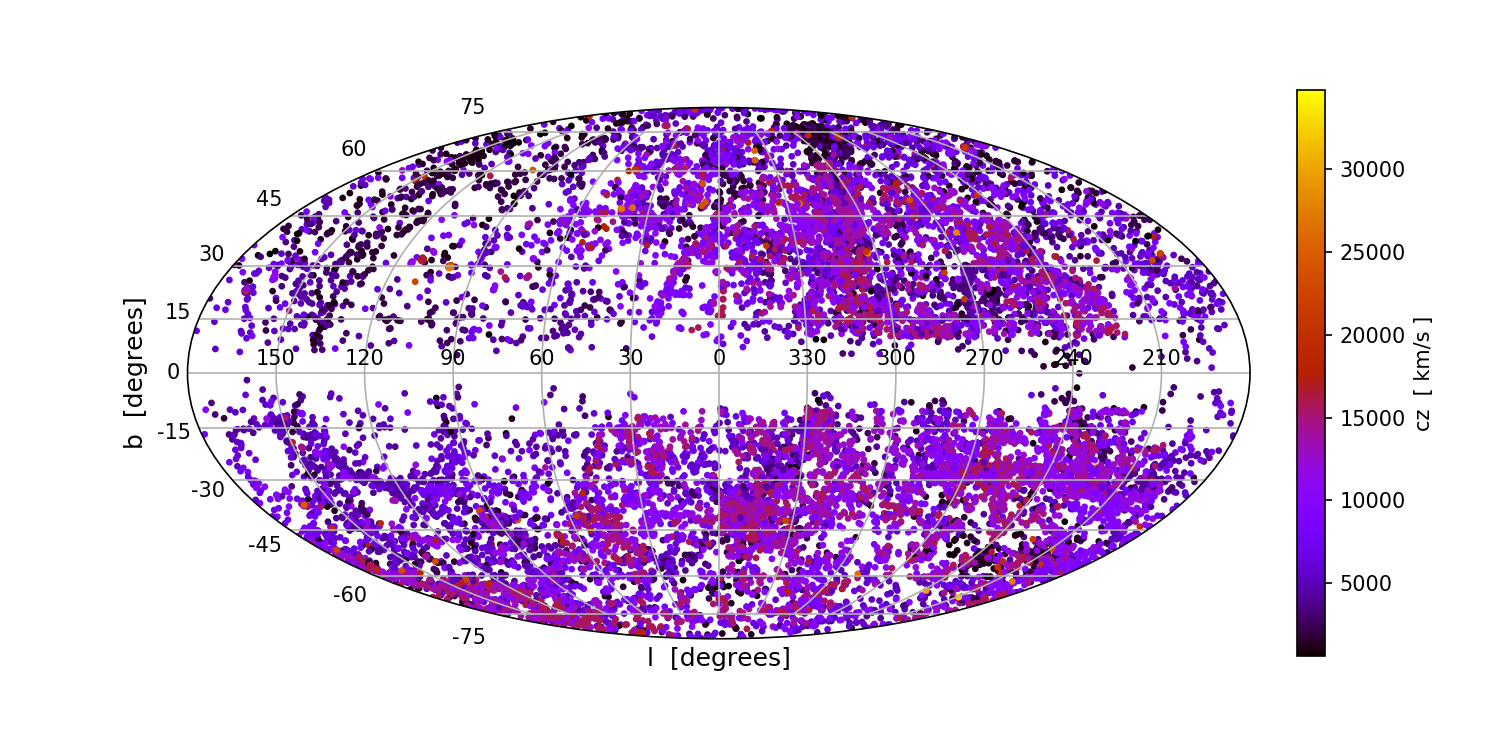}
 \caption{Distribution of 18\,352 galaxies in the combined 2MTF and CF3 dataset in Galactic coordinates. The galaxy redshift is indicated by the colour of the points, based on the right-hand colour bar. The majority of galaxies lie at recession velocities $cz<16,000$ km s$^{-1}$.}
 \label{lb}
\end{figure*}

\begin{figure} 
\centering
 \includegraphics[width=\columnwidth]{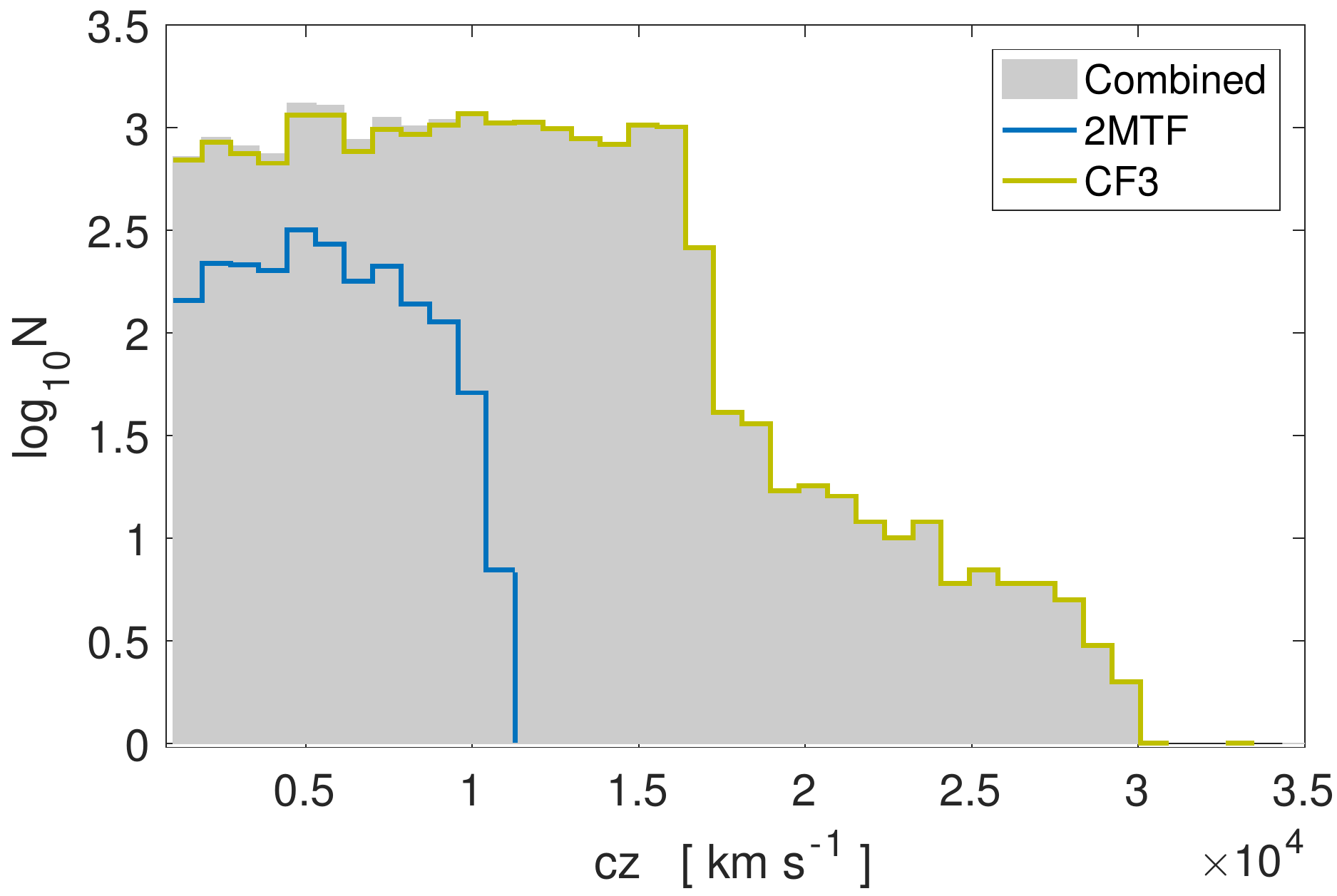}
 \caption{The redshift (in CMB frame) distribution of the datasets. The light-green/yellow and the blue line are for the CF3 and 2MTF data, respectively. The Combined dataset is represented by the gray bars. }
 \label{histcz}
\end{figure}

Removing the 1\,117 common galaxies from CF3, and adding a zero-point correction of $-0.016$ to the log-distance ratio data in CF3, we obtain a combined data set, which has 18\,352 galaxies. The sky coverage and the redshift distribution of the combined CF3 and 2MTF is shown in Fig.~\ref{lb} and Fig.~\ref{histcz}, respectively.

\section{Bulk flow and shear moments} \label{sec:BKQtheory}

Following the argument in \cite{1988MNRAS.231..149K} and \cite{1995ApJ...455...26J}, using the Taylor series expansion, we expand the line-of-sight total velocity field, $V(d_h)$ to first-order:
\be\label{vepan}
V(d_h)=B_i {\bf \hat{r}}_i+d_hA_{ij}{\bf \hat{r}}_i{\bf \hat{r}}_j+\cdot\cdot\cdot~,~~(i,j=x,y,z),
\ee
(hereafter, repeated indices represent summation), where $d_h$ is the comoving distance, and $\{{\bf \hat{r}}_x,{\bf \hat{r}}_y,{\bf \hat{r}}_z\}$ represents the projections for the unit vector of $d_h$ in the three Cartesian directions. Then, following \cite{1989MNRAS.241..787S} and \cite{2001AstL...27..765P}, we can decompose the tensor, $A_{ij}$ into a summation of trace part $H\delta_{ij}$ and traceless part $Q_{ij}$:
\be
A_{ij}=Q_{ij}+H\delta_{ij}~,~~H=\frac{1}{3}A_{ij}\delta_{ij}.
\ee
We then can write Eq.~\ref{vepan} as
\be\label{vepana}
V(d_h)-Hd_h=B_i {\bf \hat{r}}_i+d_hQ_{ij}{\bf \hat{r}}_i{\bf \hat{r}}_j+\cdot\cdot\cdot~,
\ee
where $Hd_h$ corresponds to the Hubble law with the Hubble constant $H$ \citep{2001AstL...27..765P}. 

The left hand side of Eq.~\ref{vepana}, which is the difference between the total velocity field, $V(d_h)$ and the Hubble recession velocity $Hd_h$, describes the line-of-sight peculiar motion of the galaxies. Denoting this by $v(d_h)$ gives:
\be\label{vepans}
v(d_h)=B_i {\bf \hat{r}}_i+d_hQ_{ij}{\bf \hat{r}}_i{\bf \hat{r}}_j+\cdot\cdot\cdot~.
\ee
The three zeroth-order components, $B_i$ are known as `bulk flow'. The first-order tensor, $Q_{ij}$ describes the `shear' moments and is traceless, i.e.  
\be \label{tracQ}
Q_{zz}=-Q_{xx}-Q_{yy}.
\ee
The line-of-sight peculiar velocity (PV) field only has radial components, i.e. it is curl-free. Therefore, $Q_{ij}$ is a symmetric tensor \citep{2010MNRAS.407.2328F}, $Q_{ij}=Q_{ji}$,
so that there are 5 independent shear components, giving 8 independent moment components for the first-order expansion in Eq.~\ref{vepans}.

To simplify Eq.\ref{vepans}, we follow \cite{1995ApJ...455...26J} and \cite{2010MNRAS.407.2328F} and rewrite as follows:  
\be\label{vUg}
v(d_h)=\sum^9_{p=1}U_pg_p(d_h)
\ee
where $U_p$ are the nine moment components given by
\be
U_p=\{ B_x,B_y,B_z, Q_{xx},Q_{yy},Q_{zz},Q_{xy},Q_{xz},Q_{yz} \}~,
\ee
and the mode functions are given by
\be\label{gps}
g_p(d_h)=\{
{\bf \hat{r}}_x,  
{\bf \hat{r}}_y, 
{\bf \hat{r}}_z, 
d_h{\bf \hat{r}}_x^2, 
d_h{\bf \hat{r}}_y^2,
d_h{\bf \hat{r}}_z^2,
2d_h{\bf \hat{r}}_x{\bf \hat{r}}_y, 
2d_h{\bf \hat{r}}_x{\bf \hat{r}}_z, 
2d_h{\bf \hat{r}}_y{\bf \hat{r}}_z 
\}.
\ee
In this paper, we use the measured log-distance ratio $\eta$ from the individual and combined CF3 and 2MTF samples to estimate the nine moments $U_p$.

\section{Maximum Likelihood Estimation} \label{sec:theory}

To preserve the Gaussian nature of the measurement errors, 
there are two methods that can be applied to obtain 
maximum likelihood estimates
of the bulk flow velocity and shear moments.

The first ($\eta$MLE) calculates the model log-distance ratio 
from the model $U_p$ and compares to the measured value \citep{1995MNRAS.276.1391N,2011ApJ...736...93N,2018MNRAS.477.5150Q}.

The second method ($w$MLE)  converts the measured $\eta$ into $v$-space to obtain the peculiar velocities, $v$ using the PV estimator of \cite{2015MNRAS.450.1868W}, then compares to the model $U_p$ under the assumption that the measured $v$ has Gaussian error \citep{1988MNRAS.231..149K}. 

One caveat is that the PV estimator in \cite{2015MNRAS.450.1868W} only strictly estimates an unbiased peculiar velocity under the assumption that
the $cz$ of the galaxy is much greater than the \textit{true} peculiar velocity (not the measured peculiar velocity) for that galaxy \citep{2015MNRAS.450.1868W}.  
By contrast, the $\eta$MLE can avoid assumptions about the galaxy's unknown true PV compared to its redshift.

\subsection{$\eta$MLE}\label{newmleee}

A galaxy's line-of-sight peculiar velocity can be related to its observed redshift $z$ through \citep{2001MNRAS.321..277C,2006PhRvD..73l3526H,2014MNRAS.442.1117D}
\be \label{travp}
v=c\left(\frac{z-z_h}{1+z_h}\right)~,
\ee
where redshift $z_h$ corresponds to the true comoving distance, $d_h$ of the galaxy, and $c$ is the speed of light. The above equation neglects the effects of gravitational lensing and relativistic motions \citep{2014MNRAS.442.1117D}.
In the spatially flat $\Lambda$CDM model, the comoving distance is given by
\be\label{Dz}
d_{h}(z_{h})=\frac{c}{H_0}\int_0^{z_{h}}\frac{dz'}{E(z')}\approx \frac{cz_{h}}{H_{0}}~,
\ee
where
\be\label{Ez}
E(z)=\frac{H(z)}{H_0}=\sqrt{\Omega_m(1+z)^3+\Omega_{\Lambda}}~,
\ee
and $H_{0}$, $\Omega_{m}$ and $\Omega_{\Lambda}$ are the present epoch Hubble constant, matter and dark energy densities, respectively. The apparent comoving distance $d_z$ can be related to the observed redshift $z$ through    
a similar expression.

Substituting Eq.~\ref{vepans} into Eq.~\ref{travp} to replace $v$, then using the low-redshift approximation $z_h\approx H_{0}d_{h}(z_{h})/c$ to replace $z_h$, we can obtain the relationship between $d_h$ and $\{B_i, Q_{ij}\}$: 
\begin{equation}\label{dhUp}
\begin{split}
d_h = -&\frac{cH_0+({\bf B}\cdot{\bf \hat{r}}H_0 +cQ_{ij} {\bf\hat{r}}_i  {\bf\hat{r}}_j)}{2H_0 Q_{ij} {\bf\hat{r}}_i  {\bf\hat{r}}_j }\\
+&\frac{\sqrt{({\bf B}\cdot{\bf \hat{r}}H_0+cH_0+cQ_{ij} {\bf\hat{r}}_i  {\bf\hat{r}}_j)^2+4cH_0Q_{ij} {\bf\hat{r}}_i  {\bf\hat{r}}_j(cz-{\bf B}\cdot{\bf \hat{r}})
}
}{2H_0 Q_{ij} {\bf\hat{r}}_i  {\bf\hat{r}}_j } .
\end{split}
\end{equation}
This equation is used to calculate the model-predicted $d_h$ for the $\eta$MLE and the $w$MLE. 
 The model $d_h$ is then combined with $d_z$ to compute a model log-distance ratio. 
One caveat is that, since $Q_{ij}$ in Eq.~\ref{dhUp} is traceless, the element $Q_{zz}$ is computed from $Q_{xx}$ and $Q_{yy}$ using Eq.~\ref{tracQ}, rather than setting $Q_{zz}$ as an independent shear component. 
A Taylor expansion of $d_h$ in Eq.\ref{dhUp} around the position of $(B_i=0,~Q_{ij} {\bf\hat{r}}_i  {\bf\hat{r}}_j=0)$ confirms that $d_h=d_z$, as would be expected in the absence of any peculiar velocity.

Finally, assuming that for a given set of galaxies, the measured log-distance ratios are independent and Gaussian, for a set of $n$ log-distance ratios, the likelihood can be written \citep{2018MNRAS.477.5150Q}: 
\be\label{pvpi}
P(\eta | U_p)=\prod^{N}_{n=1}\frac{1}{\sqrt{2 \pi \left(\epsilon_{n}^2+\epsilon_{\star,n}^2 \right)}}\exp\left({-\frac{1}{2}  \frac{    (\tilde{\eta}_{n}(U_p)-\eta_{n})^2  }{  \epsilon_{n}^2+\epsilon_{\star,n}^2 }}\right),
\ee
where 
$\tilde{\eta}_n(U_p)$ is the model log-distance ratio for each galaxy, $\eta_{n}$ is the measured log-distance ratio with error of
$\epsilon_{n}$, and $\epsilon_{\star,n}$ is given by  \citep{2006PhRvD..73l3526H,2014MNRAS.444.3926J}:
\be \label{errvp}
\epsilon_{\star,n}= \frac{1+z_{n}}{\mathrm{ln}(10)H(z_{n})d_{z,n}}\sigma_{\star} ,
\ee
where $\sigma_{\star}$ is the 1D velocity dispersion \citep{2016MNRAS.455..386S}. Similar to the PV estimator in \cite{2015MNRAS.450.1868W} (or our Eq.~\ref{watvp}),
Eq.~\ref{errvp} also uses the approximation that the $cz$ of the galaxy is much greater than the true peculiar velocity for that galaxy. However, in $\eta$MLE, this approximation is less important since $\sigma_{\star}$ is set to be a free parameter.

The maximum likelihood $U_p$ cannot be obtained analytically due to the non-linear  relationship between the model $U_p$ and the model predicted log-distance ratio. Instead, we follow the method of \cite{2018MNRAS.477.5150Q}, combining flat priors on the $\sigma_{\star}$ and $U_p$ (excluding $Q_{zz}$) with the likelihood in Eq.~\ref{pvpi}, enabling us to write the posterior probability of these 9 independent parameters given
the cosmological model and the data. Here, we use the Metropolis-Hastings Markov chain Monte Carlo (MCMC) algorithm with flat priors in the interval $B_{i}\in[-1200,+1200]$ km s$^{-1}$  and $Q_{ij}\in[-100,+100]$ h km s$^{-1}$ Mpc $^{-1}$ to explore the posterior.

\cite{2010MNRAS.407.2328F} use the MV method to estimate $U_p$. In their estimator, they set $Q_{zz}$ as an independent component rather than using Eq.~\ref{tracQ} to compute $Q_{zz}$ from $Q_{xx}$ and $Q_{yy}$. In our paper, we also tested the $\eta$MLE on mocks by setting $Q_{zz}$ as an independent component (see Appendix \ref{AP2}), but found this led to larger reduced $\chi^2$.

The measurement error of the bulk flow amplitude, $e_B$
can be calculated use the Jacobian, $J$ and the covariance matrix of the bulk flow velocity, $R^{\epsilon}_{ij}$ through   
\be\label{bke2}
e^2_B=JR^{\epsilon}_{ij}J^T~,~~(i=1,2,3)~,
\ee
where $J=\partial B/\partial B_i$ and $R^{\epsilon}_{ij}$ is calculated using the MCMC chains. For comparison to theory,
the `MLE depth', which is the characteristic scale of cosmic flow measurement, is defined as \citep{2016MNRAS.455..386S}
\be\label{depthri}
d_{MLE}=\frac{\sum |\boldsymbol{d}_{h,n}|W_n}{\sum W_n}~,
\ee
where the weight factors $W_n=1/(\sigma_n^2+\sigma^2_{\star})$. For the purpose of this comparison, the measurement errors of peculiar velocities, $\sigma_n$ are given by \citep{2006PhRvD..73l3526H,2014MNRAS.444.3926J,2017MNRAS.471.3135H}:  
\be \label{vperrss}
\sigma_n=\frac{\ln(10)cz_n}{1+z_n}\epsilon_n,
\ee
which is similar to Eq.~\ref{errvp}.
The theoretical expected bulk flow is compared to the measured value at the scale of $d_{MLE}$.

\subsection{\textit{w}MLE: Estimation in $v$-space} \label{sec:MLE}

Assuming peculiar velocities have Gaussian errors,
the likelihood of $n$ peculiar velocities $v_{n}$ given $U_p$ is \citep{1988MNRAS.231..149K}:
\be\label{tramle}
L(U_p,\sigma_{\star})=\prod_{n=1}^{N}\frac{1}{\sqrt{2 \pi \left(\sigma^2_n+\sigma_{\star}^2 \right)}}\exp\left(-\frac{1}{2}  \frac{    (v_{n}-\tilde{v}_n(U_p))^2  }{  \sigma^2_n+\sigma_{\star}^2 }\right)
\ee
where $\tilde{v}_n(U_p)$ is the model PV for each observed galaxy 
.

To preserve the above Gaussian assumption, \cite{2015MNRAS.450.1868W} developed the following estimator to calculate peculiar velocities as the input to the above likelihood function,
\be\label{watvp}
v=\frac{cz_{mod}}{1+z_{mod}}\ln\frac{cz_{mod}}{H_0d_l}~,~~(v_t\ll cz)
\ee
where $d_l$ is the luminosity distance, and $z_{mod}$ is given by
\be  \label{zmod}
z_{mod}=z\left[1+\frac{1}{2}(1-q_0)z-\frac{1}{6}(1-q_0-3q_0^2+1)z^2\right].
\ee
The acceleration parameter is $q_0=0.5(\Omega_m-2\Omega_{\Lambda})$. The $w$MLE method 
refers to the combination of PV estimator in Eq.~\ref{watvp} and the likelihood in Eq.~\ref{tramle}.   
However, Eq.~\ref{watvp} only estimates an unbiased PV under the assumption that the galaxy's $cz$
is much lager than its true peculiar velocity, $v_t$\citep{2015MNRAS.450.1868W}. 

To compute the peculiar velocity, we can first calculate the true comoving distance, $d_{h}$ from the measured $\eta$ and the inferred comoving distance, $d_{z}$ using
\be \label{etadist}
d_{h} = d_{z}10^{-\eta} ,
\ee
then converting to luminosity distance using $d_l=(1+z)d_h$, calculating $z_{mod}$ from the observed redshift $z$ using Eq.~\ref{zmod}. Eq.~\ref{watvp} can then be solved to obtain $v$. 
The $\tilde{v}_n(U_p)$ in Eq.~\ref{tramle} can be computed by first calculating the model-predicted $d_h$ from Eq.~\ref{dhUp}, then solving Eq.~\ref{vepans} to obtain $\tilde{v}_n(U_p)$. Similar to the $\eta$MLE method, we use MCMC with uniform priors in the interval $B_{i}\in[-1200,+1200]$ km s$^{-1}$  and $Q_{ij}\in[-100,+100]$ h km s$^{-1}$ Mpc $^{-1}$. $Q_{zz}$ is also computed from $Q_{xx}$ and $Q_{yy}$ using Eq.~\ref{tracQ}, rather than setting it as an independent component.

\section{Bulk and shear moments in the 2MTF mocks}\label{sec:mocktests}

In order to test how well the $\eta$MLE and the $w$MLE are expected to recover the true moments from the observational data, we applied the two estimators to 16 mock 2MTF catalogues \citep{2017MNRAS.471.3135H}. 
We use the SURFS simulations \citep{2018MNRAS.475.5338E} and the GiggleZ \citep{2015MNRAS.449.1454P} to generate these mocks
. The SURFS simulation uses cosmological parameters of $\Omega_m=0.3121$, $\Omega_b=0.0488$ and $h=0.6751$.
while the GiggleZ simulation uses cosmological parameters of $\Omega_m=0.273$, $\Omega_b=0.0456$ and $h=0.705$. 
This also allows us to explore whether the two estimators give consistent answers under different cosmologies. 
Each mock catalogue contains $~\sim 2\,000$ galaxies, and matches the survey geometry (i.e. sky coverage and the distance distribution) and the selection function of the 2MTF survey \citep{2018MNRAS.477.5150Q}.

The true velocity, ${\bf v}_t$ of each galaxy are known from the simulation. Within each mock, the `true' bulk flow velocity, ${\bf B}_t$ is defined as the average of the true galaxy velocities along orthogonal axes
\be
B_{t,i}=\frac{1}{N}\sum^N_{n=1}v_{t,in}~,~~(i=x,y,z)~ .
\ee
The `true' shear moments within each mock are defined as the traceless part of
\be\label{Qtij}
A_{t,ij}=\frac{1}{N}\sum^N_{n=1}\frac{v_{t,in}  {\bf \hat{r}}_{j,n}}{d_{h,n}}~ .
\ee
The true comoving distance of each galaxy, $d_{h,n}$ is known from the simulations, and ${\bf \hat{r}}_{j,n}$ is the projection, in the $j$-direction, of the corresponding unit vector.

\begin{figure}  
  \includegraphics[width=\columnwidth]{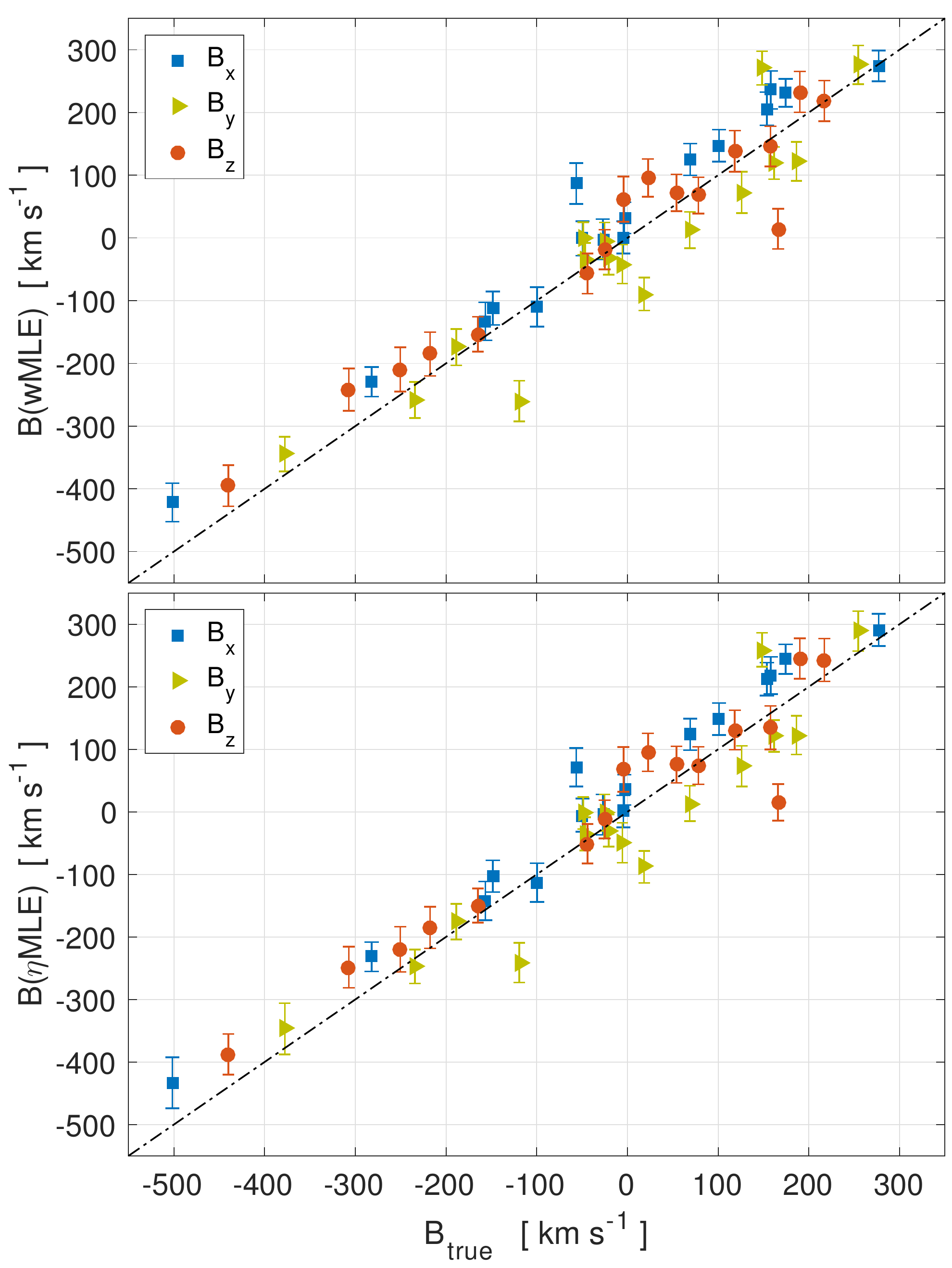}
 \caption{  The measured bulk flow for the 16 2MTF mocks in equatorial coordinates. The upper and bottom panels are for the $w$MLE estimator and the $\eta$MLE estimator, respectively.}
\label{2mtfmock}
\end{figure}

As shown in Fig.~\ref{2mtfmock}, in Cartesian equatorial coordinates, we compare the measured bulk flow of the 16 2MTF mocks to their true bulk flow. 
To compare the $\eta$MLE to $w$MLE,
we calculate the reduced $\chi^2$ between true bulk flow, ${\bf B}_{t}$ and the measured bulk flow, ${\bf B}_{m}$ using
\be\label{chi2}
\chi^2_{red}({\bf B})=\frac{1}{48-1}(\boldsymbol{B}_m-\boldsymbol{B}_{t})\boldsymbol{\mathsf{C}}^{-1}(\boldsymbol{B}_m-\boldsymbol{B}_{t})^T
\ee
where the vector ${\bf B}_{m}$ and ${\bf B}_{t}$ contain 48 elements, including 3 directions $\times$ 16 mocks. The covariance matrix $\boldsymbol{\mathsf{C}}$ contains
48$\times$48 elements (16 $3\times 3$ diagonal blocks, and zero elsewhere). The 16 $3\times 3$ diagonal blocks of $\boldsymbol{\mathsf{C}}$ are computed from the 16 MCMC samples for both $w$MLE and $\eta$MLE. For $\eta$MLE we find $\chi^2_{red}({\bf B}) = 3.70$, which is slightly lower than for $w$MLE (where $\chi^2_{red}({\bf B}) = 4.04$).
These $\chi^2_{red}({\bf B})$ are smaller compared to the results
in \cite{2018MNRAS.477.5150Q} (where $\chi^2_{red}({\bf B})=4.02$ for $\eta$MLE and 4.23 for $w$MLE).
This is because, in this work, both the $\eta$MLE estimator and the $w$MLE estimator have more parameters (due to the shear moments) which will reduce the scatter in $\boldsymbol{B}_m$ about $\boldsymbol{B}_{t}$ and increase the length of the error bars.

\begin{figure*} 
  \includegraphics[width=175mm]{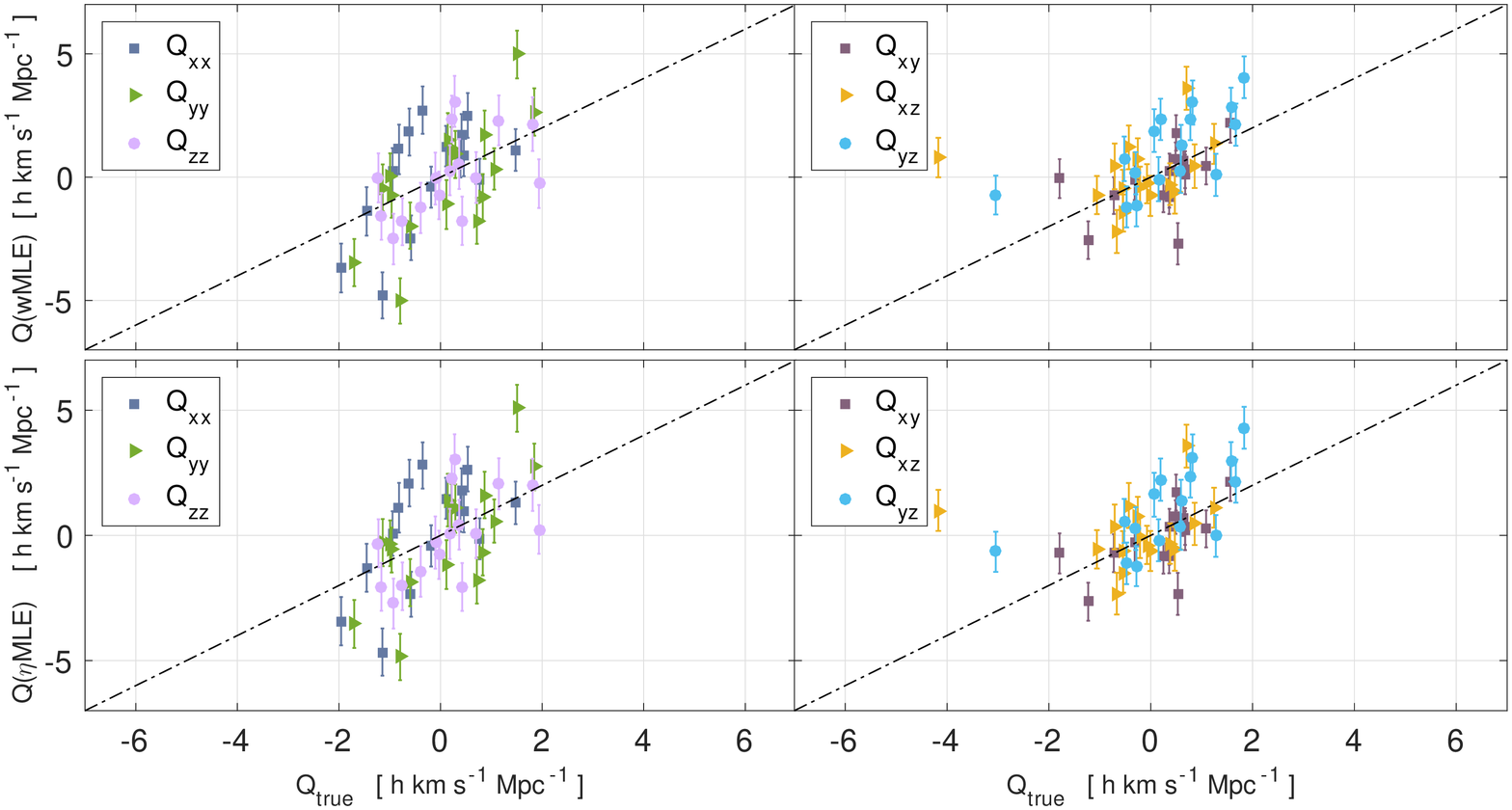}
 \caption{The measured shear moments for the 16 2MTF mocks. The $(x,y,z)$ measurements are in equatorial coordinates. The upper and lower panels are for the $w$MLE estimator and the $\eta$MLE estimator, respectively.}
\label{2mtfmockQ}
\end{figure*}

The estimated shear moments from the mocks are compared to the true shear moments in Fig.~\ref{2mtfmockQ}. Similarly, the reduced $\chi^2$ between the measured shear moments, ${\bf Q}_{m}$ and true shear moments, ${\bf Q}_{t}$ is given by
\be\label{chi2}
\chi^2_{red}({\bf Q})=\frac{1}{80-1}(\boldsymbol{Q}_m-\boldsymbol{Q}_{t})\boldsymbol{\mathsf{C}}^{-1}(\boldsymbol{Q}_m-\boldsymbol{Q}_{t})^T ,
\ee
where the ${\bf Q}_{m}$ and ${\bf Q}_{t}$ contain 80 elements (16 mocks, and 5 independent elements without $Q_{zz}$). The covariance matrix $\boldsymbol{\mathsf{C}}$ is an 80$\times$80 matrix with 16 $5\times 5$ diagonal blocks and zero elsewhere. Also, we use the 16 MCMC samples to calculate the diagonal blocks. For $\eta$MLE we find $\chi_{red}^2({\bf Q}) =3.10$, which is almost the same as the $w$MLE methods (where $\chi_{red}^2({\bf Q}) = 3.14$).

The reduced $\chi^2$ for all the 8 moments (excluding $Q_{zz}$) is given by
\be\label{chi2}
\chi^2_{red}({\bf U})=\frac{1}{128-1}(\boldsymbol{U}_m-\boldsymbol{U}_{t})\boldsymbol{\mathsf{C}}^{-1}(\boldsymbol{U}_m-\boldsymbol{U}_{t})^T ,
\ee
where the measured moments, ${\bf U}_{m}$ and the true moments, ${\bf U}_{t}$ contain 128 elements (8 independent elements and 16 mocks). The covariance matrix $\boldsymbol{\mathsf{C}}$ is an 128$\times$128 matrix with 16 $8\times 8$ diagonal blocks and zero elsewhere. The 16 MCMC samples are used to calculate the 16 $8\times 8$ diagonal blocks. For $\eta$MLE we find $\chi_{red}^2({\bf U}) =3.49$, for $w$MLE, we find $\chi_{red}^2({\bf U}) = 3.58$.

Generally, for both the bulk flow measurements and the shear measurements, the $w$MLE and $\eta$MLE perform similarly and return unbiased measurements of bulk flow and shear moments. 
However, due to the \cite{2015MNRAS.450.1868W} estimator having a necessary assumption of $v_{true}\ll cz$, some systematic errors are introduced for the closest galaxies in the mocks. 
As a result, the $\chi^2_{red}({\bf B})$ of $w$MLE is slightly higher compared to the $\eta$MLE.
Overall, we find $\eta$MLE performs better than $w$MLE for the 2MTF mocks, and for the subsequent parts of this paper, $\eta$MLE is the one we shall adopt to measure the bulk flow and shear moments from the datasets.

The reasons for the reduced chi-squared values far from 1 are most likely due to: (a) the assumption that the standard deviation of true velocities in the mocks is $\sigma_{\star}$ (or $\epsilon_{\star,n}$ in the $\eta$MLE method); (b) the fitted values of bulk and shear flow are weighted in a different manner to the `true' values, leading to different effective depths; and (c) the moment model is only a low-order approximation of a more complex velocity field.

\section{Results and discussion}\label{sec:dis}

\subsection{Results and comparison with $\Lambda$CDM theory}\label{sectqqq}

The resultant bulk and shear moments measurements (in Galactic coordinates) for 2MTF, CF3 and the combined data are presented in Table~\ref{bkflb}. The measurement errors of the bulk flow velocity and shear moments for the combined dataset (and CF3) are much smaller compared to 2MTF. This is mainly due to the combined dataset covering a much larger cosmological volume. CF3 also combines distances using the weighted average of multiple measurements, if available.

\begin{figure*}  
  \includegraphics[width=175mm]{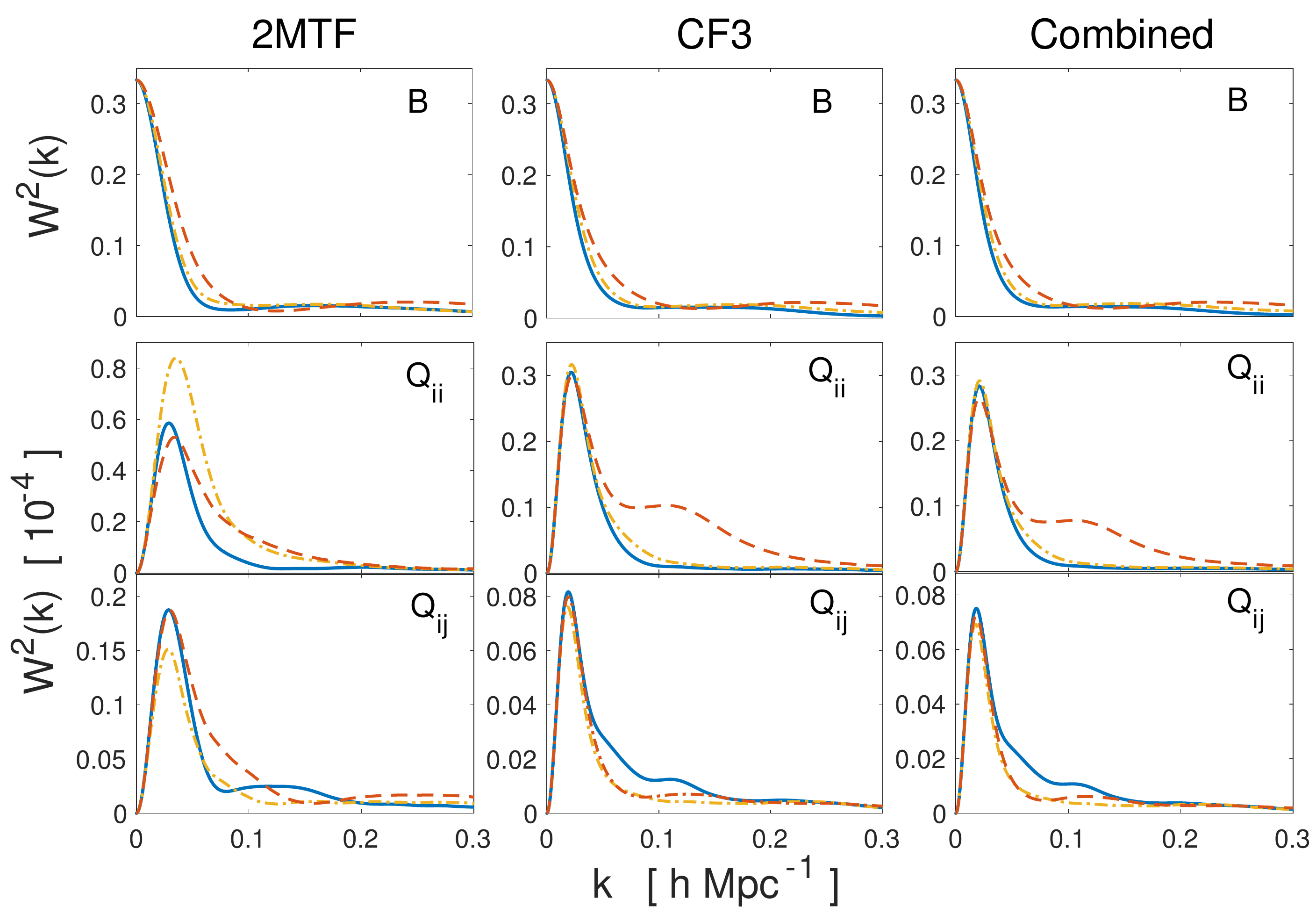}
 \caption{The window functions for 2MTF, CF3 and the combined dataset, all in Galactic coordinates. In the top panels are the bulk flow $x$ (blue solid curve), $y$ (yellow dot-dashed curve) and $z$ (red dashed curve) components. In the middle panels are the $Q_{xx}$ (blue solid curve), $Q_{yy}$ (yellow dot-dashed curve), $Q_{zz}$ (red dashed curve) components. In the bottom panels are the $Q_{xy}$ (blue solid curve), $Q_{xz}$ (yellow dot-dashed curve), $Q_{yz}$ (red dashed curve) components. The left-hand side panels are for 2MTF, the middle panels are for CF3, the right-hand side panels are for the combined dataset.}
\label{WFunc}
\end{figure*}

In Table~\ref{bkflb}, we also list 
the $\Lambda$CDM prediction, which has zero mean and `cosmic root mean square' (CRMS) variation \citep{2010MNRAS.407.2328F}, for each dataset.   
Assuming $\Lambda$CDM, the CRMS is given by the diagonal elements of the following covariance matrix \citep{2010MNRAS.407.2328F,2011PhRvD..83j3002M,2014MNRAS.444.3926J}:
\be \label{Rcrms}
R_{pq}^{v}=\frac{\Omega^{1.1}_mH^2_0}{2\pi^2} \int\mathcal{W}_{pq}^2(k)\mathcal{P}(k)dk~.
\ee
The indices $p$ and $q$ range from 1 to 9, corresponding to the 9 moment modes in Eq.~\ref{gps}.  The linear density power spectrum $\mathcal{P}(k)$ is generated using the {\tiny CAMB} package \citep{Lewis:1999bs}. The window function, $\mathcal{W}_{pq}(k)$ for the individual moments $U_p$ is given by \citep{2010MNRAS.407.2328F,2011PhRvD..83j3002M,2014MNRAS.444.3926J}:
\be\label{wdfcs}
\mathcal{W}_{pq}^2(k)=\sum_{m,n}^Nw_{p,m}w_{p,n}f_{mn}(k) ~.
\ee
The analytic expression of the angle-averaged window function, $f_{mn}(k)$ is given by \cite{2011PhRvD..83j3002M} (also, see Equation 5 in \citealt{2014MNRAS.444.3926J}). 
Assuming PVs have a Gaussian distribution (see Eq.~\ref{tramle}), the weight factors, $w_{p,n}$ in Eq.~\ref{wdfcs} are given by \citep{1988MNRAS.231..149K,1995ApJ...455...26J,2008MNRAS.387..825F}:
 \be \label{aijml3}
w_{p,n}=\sum^9_{q=1}A_{pq}^{-1}\frac{g_{q,n}}{\sigma^2_n+\sigma^2_{\star}}~,~~
A_{pq}=\sum^N_{n=1}\frac{g_{p,n}g_{q,n}}{\sigma^2_n+\sigma^2_{\star}}~.
\ee
The above Gaussian assumption is true for the $w$MLE. Since the $w$MLE and the $\eta$MLE give almost the same $U_p$ measurements (or almost the same $\chi^2_{red}$, see Fig.~\ref{2mtfmock}), it is rational to compare the `CRMS', inferred from Eq.~\ref{aijml3}, to the $\eta$MLE measurements in Table~\ref{bkflb}, even though the weight factors correspond to velocities rather than log-distance ratios. In Galactic coordinates, the $\mathcal{W}_{pq}^2(k)$ of Eq.~\ref{wdfcs} for 2MTF, CF3 and the combined datasets is shown in Fig.~\ref{WFunc}.

The estimation procedure of CMRS is follows the arguments in \cite{2010MNRAS.407.2328F} (see also \citealt{2009MNRAS.392..743W,2016MNRAS.455..386S}).
To reiterate, to obtain the estimates of the CMRS expected within our survey under the $\Lambda$CDM cosmological model, we perform the following steps:
\begin{enumerate}
\item{Use the positions and errors of galaxies within the 2MTF (CF3) survey to calculate the weight factors in Eq.~\ref{aijml3}.}
\item{Combine these with the $f_{mn}$ term in Eq.~\ref{wdfcs} (which also only depends on the 2MTF (CF3) galaxy positions) to calculate the window function of the data.}
\item{Integrate this window function along with the $\Lambda$CDM power spectrum to calculate the matrix $R_{pq}$ in Eq.~\ref{Rcrms}. }
\item{The CMRS values are then given by the diagonal elements of $R_{pq}$.}
\end{enumerate}
As the above procedure depends only on the positions and errors of the galaxies within 2MTF (CF3) data and the $\Lambda$CDM cosmological model, we are testing the agreement between our data and the cosmological model without the need for any simulations.

From Table~\ref{bkflb}, for the CF3, 2MTF and the combined sample, we find that the majority of our $Q_{ij}$ measurements are consistent with the CRMS calculated from $\Lambda$CDM theory. After combining the CRMS predictions with measurement errors, the largest deviations are the $Q_{xy}$ component in 2MTF at 2.2$\sigma$ and the $Q_{zz}$ component in the combined dataset at 2.2$\sigma$ too.   

We also need to compare the measured bulk flow amplitude to the $\Lambda$CDM prediction. Unlike the individual components of the bulk flow and shear, the bulk flow \textit{amplitude} is non-Gaussian, following instead a Maxwell-Boltzmann distribution. The rms of the bulk flow amplitude, $\sigma_B$ can be calculated from $R_{ij}^{v}$ of Eq.~\ref{Rcrms} using the Jacobian.
The probability distribution of the bulk flow amplitude $B$ is given by \citep{2012ApJ...761..151L,2014MNRAS.445..402H,2016MNRAS.455..386S}
\be\label{pbb}
p(B)=\sqrt{\frac{2}{\pi}}\left(\frac{3}{\sigma_B^2}\right)^{1.5}B^2\exp\left(-\frac{3B^2}{2\sigma_B^2}\right)
\ee
where the most likely $B$ is expressed as $B_p=\sqrt{2/3}\sigma_B$, and the cosmic variance of B 
is given by $B_p$$^{+0.419\sigma_B}_{-0.356\sigma_B}$ (68\% confidence level) and $B_p$$^{+0.891\sigma_B}_{-0.619\sigma_B}$ (95\% confidence level) \footnote{The upper and lower limits mean that the integral of Eq.\ref{pbb} in the interval $\left[ B_p-0.356\sigma_B,  B_p+0.419\sigma_B   \right]$ is 0.68. The integral in the interval $\left[ B_p-0.619\sigma_B,  B_p+0.891\sigma_B   \right]$ is 0.95. The interesting question is if we were to calculate the bulk flow around $N$ random $\Lambda$CDM observers, what would be the expected value of the bulk flow (the answer is $B_{p}$) and where would 68\% (95\%) of the measurements lie about this point. Then comparing this statistic to our measured local bulk flow as a test of whether or not our measurement would be expected within a $\Lambda$CDM universe.}  \citep{2016MNRAS.455..386S}.
The theoretical bulk flow amplitude prediction for the 2MTF, CF3 and combined dataset is given in Table \ref{bkvst}. All bulk flows are consistent with $\Lambda$CDM predictions.

\begin{table*}   \small
\caption{Bulk flow and shear moments measurements for 2MTF, CF3 and the combined dataset in Galactic coordinates.  The CRMS column gives expected cosmic variance due to $\Lambda$CDM. The last row lists the depth of the measurement.}
\begin{tabular}{|c|c|c|c|c|c|c|}
\hline
\hline
      & \multicolumn{2}{|c|}{2MTF} & \multicolumn{2}{|c|}{CF3}& \multicolumn{2}{|c|}{Combined}  \\
\hline
     & $\eta$MLE & CRMS  & $\eta$MLE & CRMS& $\eta$MLE & CRMS\\
\hline
  $B_x$ (km s$^{-1}$)  & $130.6\pm39.5$ & $\pm$164.8  & $134.5\pm17.4$ & $\pm$160.4& $120.6\pm17.7$ & $\pm$155.6\\
\hline
  $B_y$ (km s$^{-1}$)  & $-340.3\pm37.0$ & $\pm$172.1  & $-282.7 \pm15.7$& $\pm$169.4& $-206.5\pm16.1$ & $\pm$164.2\\
\hline
  $B_z$ (km s$^{-1}$)  & $85.4\pm30.5$ & $\pm$185.6  & $76.6\pm11.9$ & $\pm$178.3& $99.5\pm12.1$ & $\pm$173.4\\
\hline
  $Q_{xx} $  ($h$ km s$^{-1}$ Mpc$^{-1}$)   & $3.69\pm1.35$ & $\pm$2.53  & $0.73\pm0.43$ & $\pm$1.78& $2.12\pm0.44$ & $\pm$1.68\\
\hline
  $Q_{xy} $  ($h$ km s$^{-1}$ Mpc$^{-1}$)  & $-4.16\pm1.16$ & $\pm$1.45  & $-1.41\pm0.36$ & $\pm$0.95& $-1.16\pm0.36$ & $\pm$0.89\\
\hline
  $Q_{xz} $  ($h$ km s$^{-1}$ Mpc$^{-1}$)   & $-0.96\pm0.97$ & $\pm$1.32  & $1.13\pm0.28$ & $\pm$0.86& $0.17\pm0.27$ & $\pm$0.80\\
\hline
  $Q_{yy} $  ($h$ km s$^{-1}$ Mpc$^{-1}$)  & $0.36\pm1.30$ & $\pm$3.20  & $0.83\pm0.41$ & $\pm$1.84& $1.94\pm0.41$ & $\pm$1.74\\
\hline
  $Q_{yz}$($h$ km s$^{-1}$ Mpc$^{-1}$)   & $-1.47\pm1.01$ & $\pm$1.54  & $-0.14\pm0.29$ & $\pm$0.89& $0.47\pm0.29$ & $\pm$0.83\\
 \hline
  $Q_{zz} $  ($h$ km s$^{-1}$ Mpc$^{-1}$)   & $-4.05\pm1.18$ & $\pm$2.68   & $-1.56\pm0.36$ & $\pm$2.02& $-4.06\pm0.36$ & $\pm$1.85\\
 \hline
$d_{MLE}$   ($h^{-1}$ Mpc) & \multicolumn{2}{|c|}{32} & \multicolumn{2}{|c|}{35}& \multicolumn{2}{|c|}{37}  \\  
      \hline
   
\end{tabular}
 \label{bkflb}
\end{table*}

\begin{table}   \centering
\caption{The $\eta$MLE measured bulk flows are compared to the prediction of $\Lambda$CDM and its cosmic variance.}
\begin{tabular}{|c|c|c|}
\hline
\hline
&$\Lambda $CDM |{\bf B}| (km s$^{-1}$)&$\eta$MLE |{\bf B}| (km s$^{-1}$)\\
\hline
CF3   & 238$^{+122}_{-104}$ & $322\pm15$  \\
\\
2MTF   & 243$^{+125}_{-106}$   & $374\pm36$ \\
\\
Combined & 231$^{+118}_{-101}$ & $259\pm15$ \\

\hline
\end{tabular}
 \label{bkvst}
\end{table}

\subsection{Cosmic flow as a function of depth}

\begin{figure*} 
\includegraphics[width=170mm]{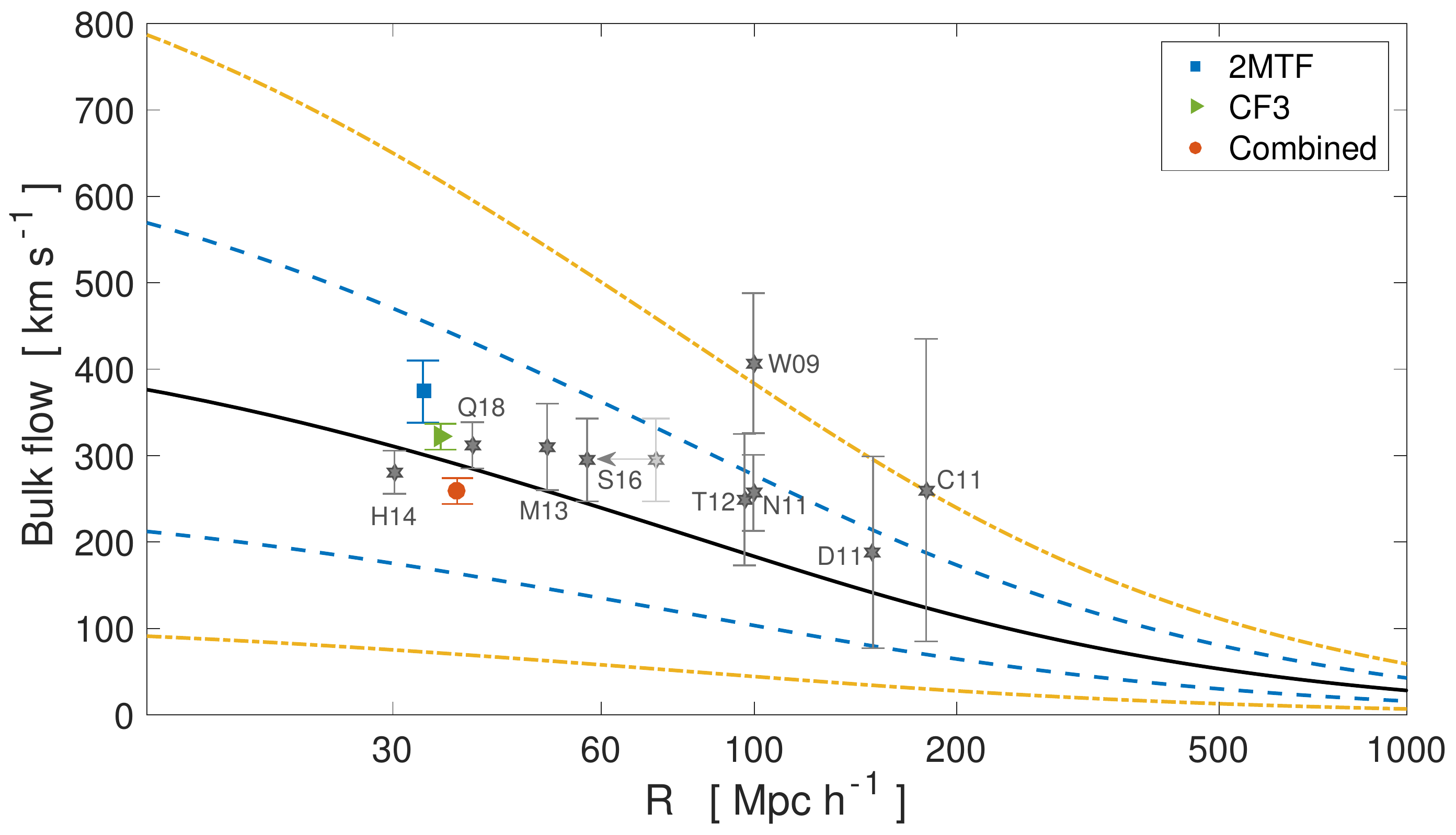}
\caption{The $\eta$MLE measured bulk flow amplitudes (filled circles $\bullet$) for 2MTF, CF3 and the combined dataset are compared to values from other authors. The most probable bulk flow from the $\Lambda$CDM prediction is shown as the solid line. 
The yellow and blue dashed lines indicate 95\% and 68\% confidence levels, respectively.   
 Other measurements are indicated by the grey stars ($\star$) 
(Q18: \citet{2018MNRAS.477.5150Q};
 H14: \citet{2014MNRAS.445..402H}; 
 T12: \citet{2012MNRAS.420..447T};
 W09: \citet{2009MNRAS.392..743W};  
 N11: \citet{2011ApJ...736...93N}; 
 M13: \citet{2013MNRAS.428.2017M}; 
 D11: \citet{2011JCAP...04..015D};
 C11: \citet{2011MNRAS.414..264C}; 
 S16: \citet{2016MNRAS.455..386S}). Following \citet{2016MNRAS.455..386S}, W09 and T12 are plotted at twice their quoted radius since they
use Gaussian windows. To account for the half-sky coverage of 6dFGSv, the S16 \citep{2016MNRAS.455..386S} measurement is shifted as shown by the grey arrow.}
\label{BvsR} 
\end{figure*}

The bulk flow amplitudes, measured from 2MTF and the CF3 individually and combined, are plotted against the survey depth in Fig.~\ref{BvsR}. 
Usually, comparing bulk flow measurements between different surveys on a single figure is difficult, since those surveys have differing survey geometries and depths. Therefore, it is necessary to standardise the window function, and in this paper, we used the spherical top-hat window function: $\mathcal{W}(k)=3(\sin kR-kR\cos kR)/(kR)^3$. In Fig.~\ref{BvsR}, we also compare our bulk flow measurements with the measurements of others \citep{ 2009MNRAS.392..743W,2011MNRAS.414..264C,2011JCAP...04..015D,2011ApJ...736...93N,2012MNRAS.420..447T,2013MNRAS.428.2017M,2014MNRAS.445..402H,2016MNRAS.455..386S,2018MNRAS.477.5150Q}. The black solid curve represents the most likely bulk flow predicted by the $\Lambda$CDM using the spherical top-hat window function.
From Fig.~\ref{BvsR}, we find most of the measured bulk flows are consistent with the $\Lambda$CDM prediction at the 68\% Confidence level.

\begin{figure*} 
\includegraphics[width=175mm]{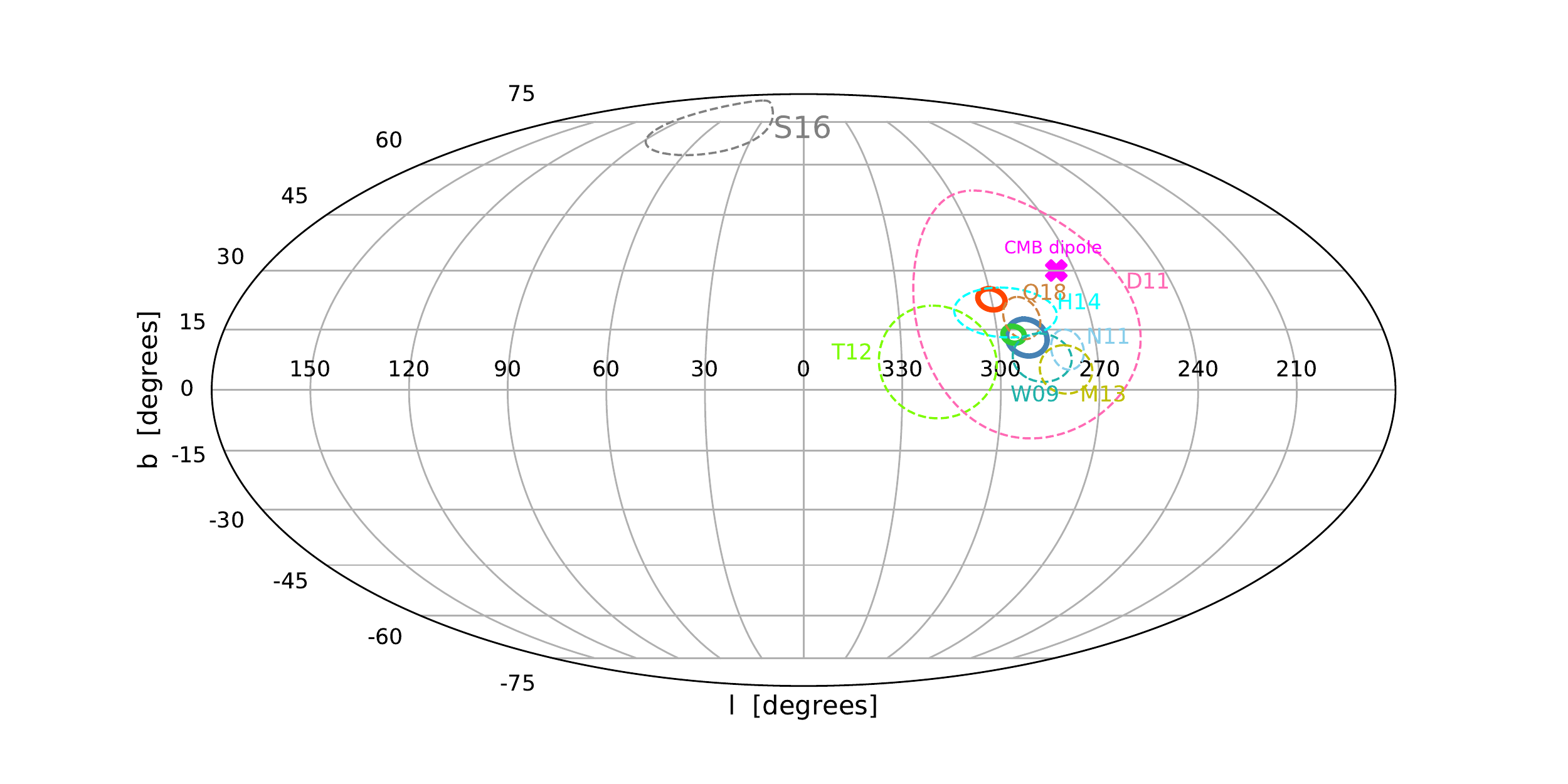}
\caption{The direction of bulk flow from different measurements are compared in Galactic coordinates. The $\eta$MLE results for 2MTF, CF3 and the combined dataset are shown in
the blue, green and the red solid circles, respectively.
The coloured dashed circles indicate
other recent measurements 
(Q18:\citet{2018MNRAS.477.5150Q};
 H14: \citet{2014MNRAS.445..402H}; 
 T12: \citet{2012MNRAS.420..447T};
 W09: \citet{2009MNRAS.392..743W};  
 N11: \citet{2011ApJ...736...93N}; 
 M13: \citet{2013MNRAS.428.2017M}; 
 D11: \citet{2011JCAP...04..015D};
 S16: \citet{2016MNRAS.455..386S}). The $1\sigma$ error is indicated by the radius of the circles. The CMB dipole direction is shown as the pink cross.}
\label{lbcomp} 
\end{figure*}

In Fig.~\ref{lbcomp}, the bulk flow directions are compared in Galactic coordinates. 
The bulk flow directions from different surveys
are mainly in agreement except S16 \citep{2016MNRAS.455..386S}. This discrepancy appears to come from their imperfect Malmquist bias correction to the 6dFGSv data, based on the assumption that peculiar velocities, estimated from Eq.~\ref{travp}, have Gaussian errors (see \citealt{2018MNRAS.477.5150Q}). The bulk flow direction converges towards the CMB dipole, and appears to be due to local effects combined with more distant gravitational perturbations, including the Shapley supercluster.

\begin{figure} 
\includegraphics[width=\columnwidth]{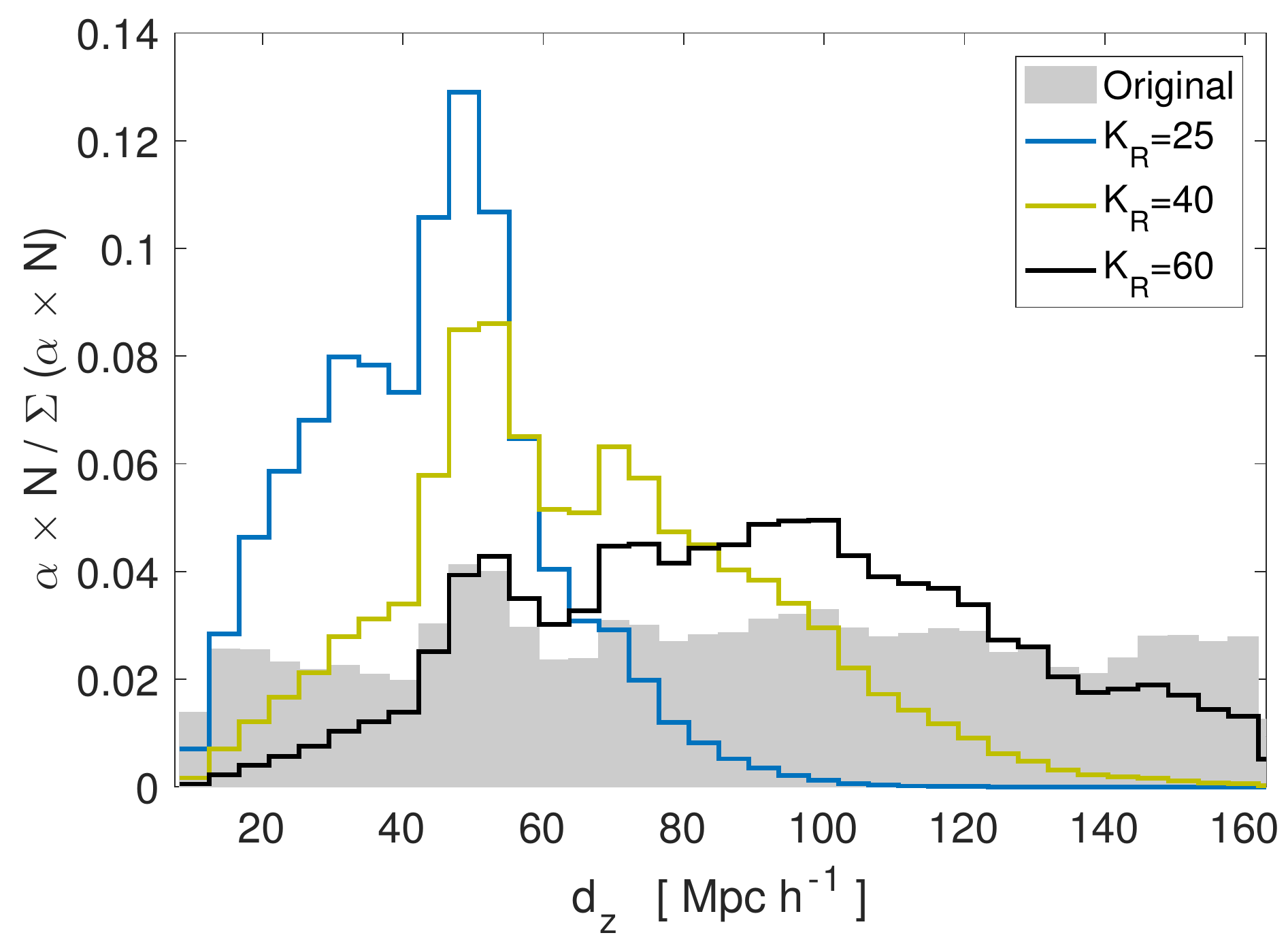}
\caption{The distribution of $d_z$ setting $K_R=25$, $40$, and $60$ respectively. The gray bars are  the original distribution of $d_z$ without any weighting.}
\label{weidz} 
\end{figure}

\begin{figure} 
\includegraphics[width=\columnwidth]{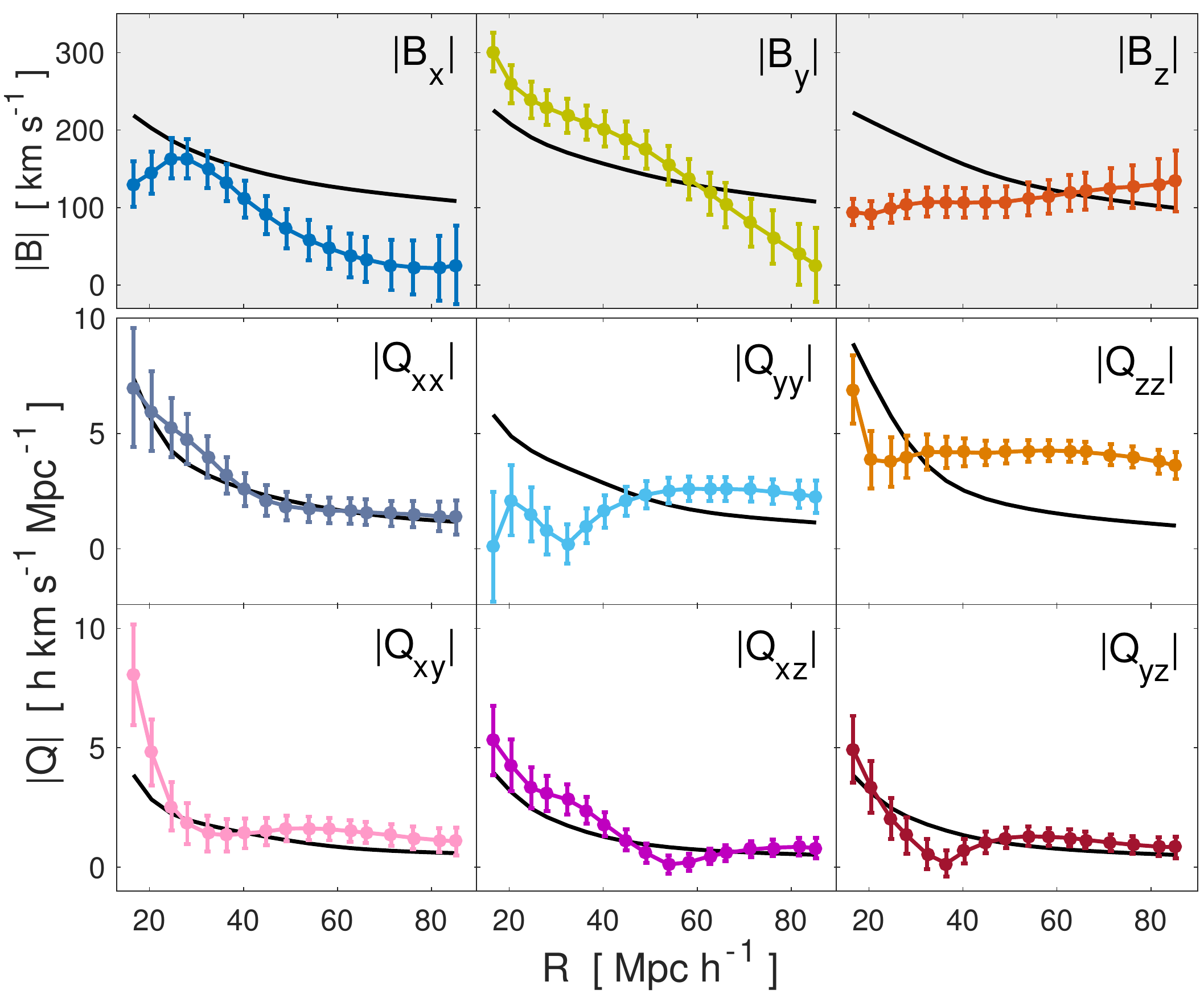}
\caption{The absolute amplitude of the moments as a function of survey depth. The upper panels are for the bulk flow, the middle panels are for the diagonal elements of the shear tensor, and the bottom panels are for the non-diagonal elements of the shear tensor. The black solid curves are the $\Lambda$CDM CRMS predictions for each moment. The measurement points for the components are highly correlated, so their covariance must be taken into account when comparing to the black curves. }
\label{QijvsRwei} 
\end{figure}

We can also use the data to explore how the measured and theoretical moments compare at different depths. By changing galaxy's contribution to the likelihood of Eq.~\ref{pvpi}, the measured bulk and shear moments will change along with the survey depth. Using the combined dataset, 
in order to adjust each
galaxy's contribution to Eq.~\ref{pvpi}, we multiply the \textit{logarithmic} likelihood of the $n$-th galaxy by the following  weight factors
\be \label{ddkks}
\alpha_n=\frac{K_n}{max(K_n)}~,~where~~K_n=d_{z,n}^2\exp\left(-\frac{d_{z,n}^2}{2 K_R^2}\right)~.
\ee
By changing the value of $K_R$,
we can change the distribution of $d_z$, as shown in Fig.~\ref{weidz}. Given $K_R$, the survey depth is calculated from a modified version of Eq.~\ref{depthri}: 
\be\label{depthri2}
R=\frac{\sum |\boldsymbol{d}_{h,n}|W_n\alpha_n}{\sum W_n\alpha_n}~.
\ee
In Fig.\ref{QijvsRwei}, we plot the measured \textit{absolute} amplitudes of the moments against $R$.  
The black solid curves are the  CRMS for each of the moments generated using the combined data set by multiplying Eq.~\ref{aijml3} 
by the weight factors, $\alpha_n$:
\be
w_{p,n}=\sum^9_{q=1}A_{pq}^{-1}\frac{\alpha_ng_{q,n}}{\sigma^2_n+\sigma^2_{\star}}~,~~
A_{pq}=\sum^N_{n=1}\frac{\alpha_ng_{p,n}g_{q,n}}{\sigma^2_n+\sigma^2_{\star}}~.
\ee
then repeating the steps given in Section \ref{sectqqq}.
As shown in Table \ref{chi2tab3}, corresponding to Fig.~\ref{QijvsRwei},
 we list the $\chi^2$ difference between the measured and the theoretical moments, and the probability of obtaining a larger $\chi^2$, $P(>\chi^2)$ at different depths. In all cases we do not find sufficient evidence to reject $\Lambda$CDM with average level of 40 per cent.

\begin{table}   \small  \centering
\caption{The $\chi^2$ and probability $P(>\chi^2)$ of the measured moments at different depths.  The degrees of freedom for ${\bf B}$, ${\bf Q}$ and ${\bf U}$ are 3, 5 and 8, respectively.}
\begin{tabular}{|c|c|c|c|c|c|c|}
\hline
\hline
  $R$     & \multicolumn{2}{|c|}{{\bf B}} & \multicolumn{2}{|c|}{{\bf Q}}& \multicolumn{2}{|c|}{{\bf U}}  \\
   $h^{-1}$ Mpc  & $\chi^2$     &   $P$     & $\chi^2$&    $P$   & $\chi^2$&      $P$   \\
\hline
   20   & 2.284 &  0.52  & 7.051 &  0.22  & 9.0160 &  0.34     \\    
\hline
   28   & 2.793 &  0.42  & 4.774 &  0.44  & 6.7161 &  0.57   \\
\hline
   36   & 2.761 &  0.43  & 3.989 &  0.55  & 6.2778 &  0.62   \\      
\hline
   45   & 2.454 &  0.48  & 3.778 &  0.58  & 6.2926 &  0.61   \\
\hline
   54   & 2.123 &  0.55  & 5.204 &  0.39  & 8.1539 &  0.42   \\      
\hline
   58   & 1.979 &  0.58  & 6.083 &  0.30  & 9.3231 & 0.32   \\      
\hline
   63   & 1.882 &  0.60  & 6.869 &  0.23  & 10.3767 & 0.24      \\      
\hline
   66   & 1.803 &  0.61  & 7.327 &  0.20  & 10.9342  & 0.21      \\      
\hline
   71   & 1.726 &  0.63  & 7.780 &  0.17  & 11.4380 & 0.18    \\ 
\hline
   76   & 1.671 &  0.64  & 7.804 &  0.17  & 11.1989 & 0.19     \\  
\hline
   81   & 1.675 &  0.64  & 7.685 &  0.17 & 10.7017 &  0.22    \\  
\hline
   85   & 1.739 &  0.63  & 7.029 &  0.22 & 9.7784 &  0.28    \\
\hline
\end{tabular}
 \label{chi2tab3}
\end{table}

\section{Conclusions}\label{conc}

We have measured the bulk and shear moments in the individual and combined 2MTF and CF3 surveys. We applied the $\eta$MLE to the catalogues in order to preserve the Gaussian nature of the measurement errors of the peculiar velocities.
Using the galaxies common between 2MTF and CF3, we demonstrate a small zero-point difference of $-0.016\pm0.002$ dex. 

We have tested the $\eta$MLE on 2MTF mocks and compare to the $w$MLE results. We find $\eta$MLE performs better than $w$MLE in both the bulk and shear moment estimation. In addition, by performing tests on anisotropic mocks, we found that leaving $Q_{zz}$ (or the trace of shear tensor) as a free parameter in the MCMC routine of $\eta$MLE is not desirable, and increases the measurement error of $Q_{zz}$ significantly.

We compare the measured bulk and shear components to the predictions from $\Lambda$CDM model
and the measurements to be consistent with the $\Lambda$CDM prediction, with no substantial deviation from the cosmic RMS values predicted by $\Lambda$CDM. Using the combined dataset, we have also explored the change of bulk and shear moments  with survey depth and again find consistency with $\Lambda$CDM at all depths between 20 and 85 Mpc $h^{-1}$.  
Using the combined sample, we measured the amplitude (depth) of the bulk flow to be $259\pm15$ km s$^{-1}$ ($37h^{-1}$ Mpc), the result again being consistent with the $\Lambda$CDM prediction at the 68\% confidence level.

\section*{Acknowledgements}
Fei Qin has received financial support from China Scholarship Council (CSC). Tao Hong is supported by the Open Project Program of the Key Laboratory of FAST, NAOC, Chinese Academy of Sciences. Parts of this research were conducted by the Australian Research Council Centre of Excellence for All-sky Astrophysics (CAASTRO), through project number CE110001020 and the Australian Research Council Centre of Excellence for All Sky Astrophysics in 3 Dimensions (ASTRO 3D), through project number CE170100013.



\bibliographystyle{mnras}
\bibliography{BKFmn}



\appendix

\section{setting $Q_{\lowercase{zz}}$ as an independent parameter in $\eta$MLE}\label{AP2}

\begin{figure}  
\includegraphics[width=\columnwidth]{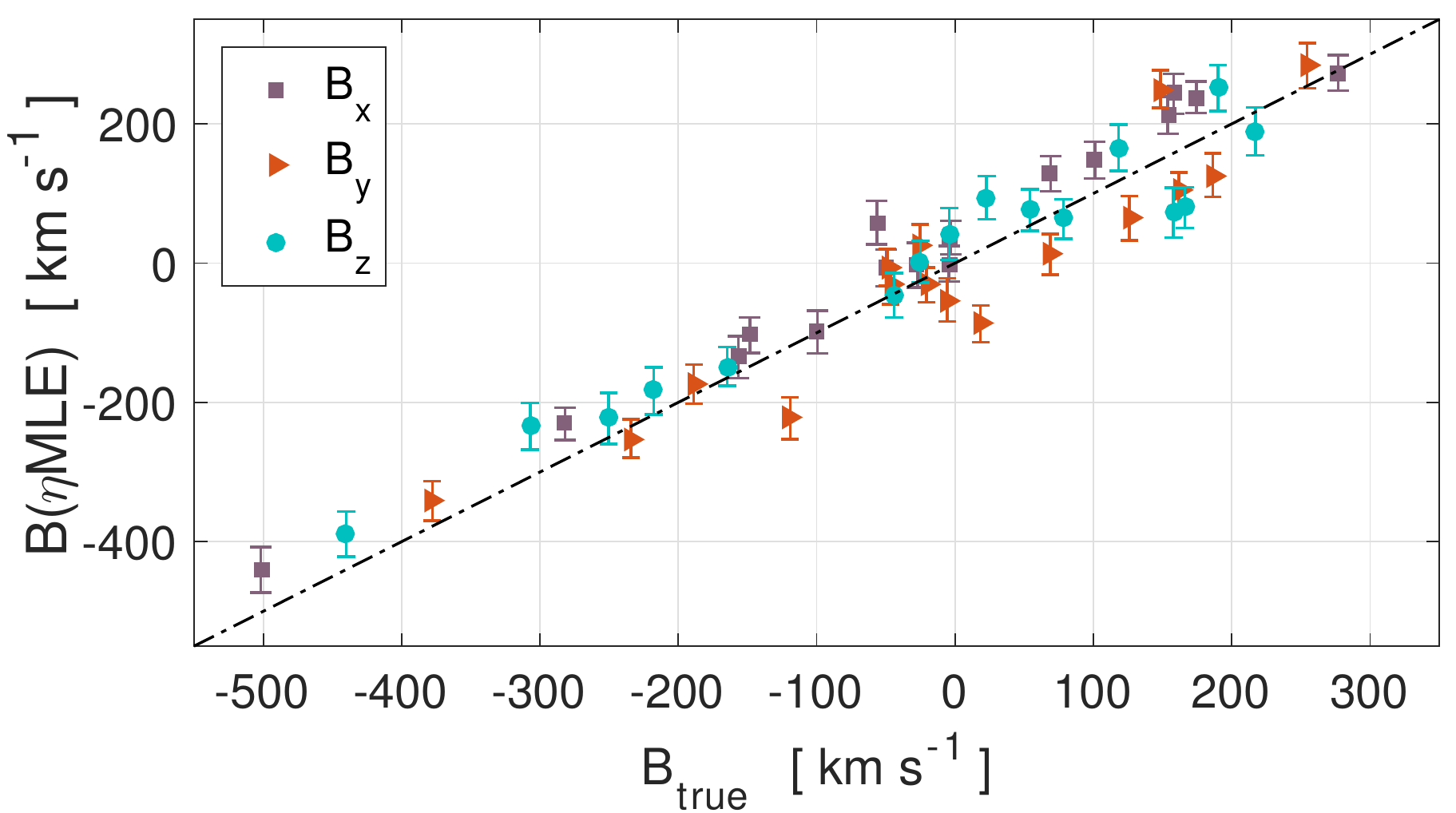}
\includegraphics[width=\columnwidth]{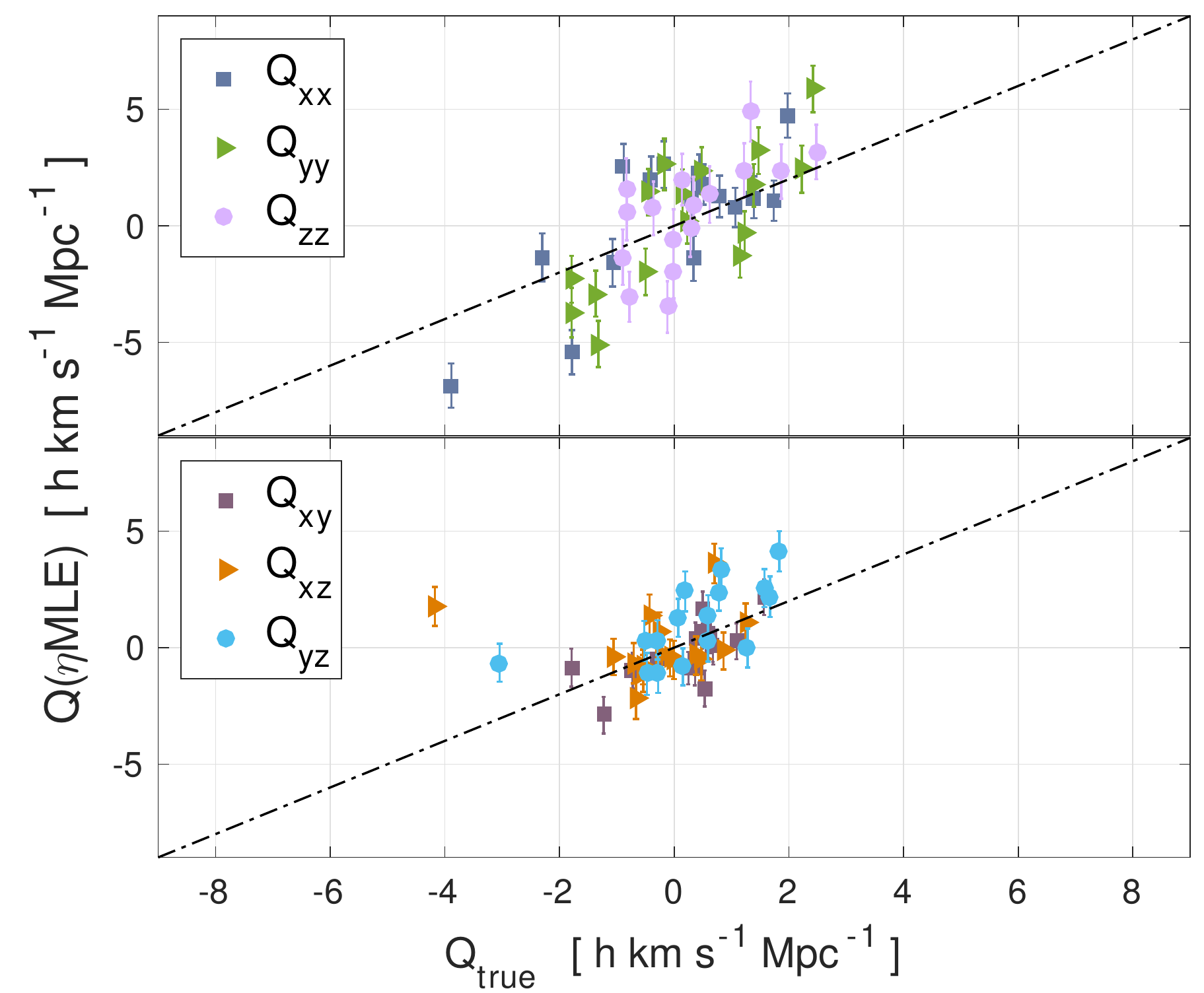}
 \caption{The measured bulk flow and shear for 16 2MTF mocks in equatorial coordinates. $Q_{zz}$ is set to be an independent component in the MCMC routine of the estimators. The top panel is for the bulk flow measurements; the middle and bottom panels are for the shear measurements.}
\label{2mtfmocktrB}
\end{figure}

In the MCMC routine of $\eta$MLE, by setting $Q_{zz}$ as an independent shear component,  
 we measured the bulk flow for the 16 2MTF mocks in equatorial coordinates, and compare to the true bulk flow in the top panel of Fig.~\ref{2mtfmocktrB}. We find $\chi^2_{red}({\bf B}) = 3.73$. 
The measured shear moments from the mocks are shown in middle and bottom panels of Fig.~\ref{2mtfmocktrB}. Correspondingly, the true shear moments is calculated directly from Eq.~\ref{Qtij} without removing the trace. The $\chi^2_{red}({\bf Q})$ is $3.63$.
For all the 8 moments,  we find $\chi^2_{red}({\bf U})$ is $3.80$.

Compared with the $\chi^2_{red}$ values from $\eta$MLE in Section \ref{sec:mocktests},
 we find that setting $Q_{zz}$ as an independent component in the MCMC routine, results in larger $\chi^2_{red}$ values. There therefore appears to be no gain in setting $Q_{zz}$ as an independent component.

\begin{figure} 
 \includegraphics[width=\columnwidth]{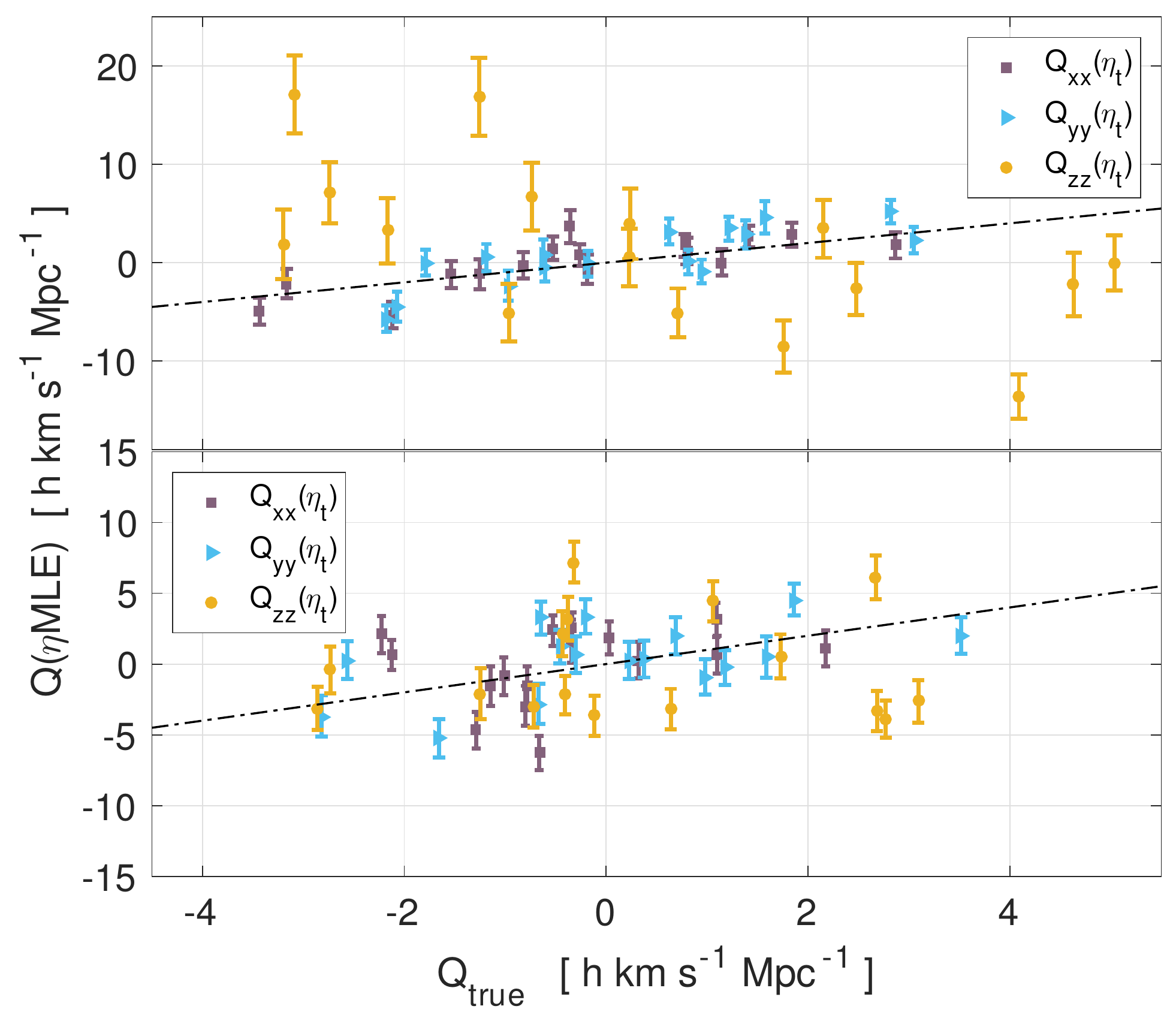}
 \caption{The measurements of the diagonal elements of ${\bf Q}$ for the half-sky 2MTF mocks using $\eta$MLE. In the upper panel, we set $Q_{zz}$ as an independent component in the MCMC routine. In the bottom panel, $Q_{zz}$ is not independent in the MCMC routine.}
\label{6dfmockqii}
\end{figure}

For anisotropic sky coverage, setting $Q_{zz}$ as an independent component in the MCMC routine of $\eta$MLE results in worse biases from the true values. As an example, we removed mock galaxies in the northern sky (Dec>0$^{\circ}$) to obtain a half-sky 2MTF mocks. Then we used the true log-distance ratio, $\eta_t$ to measure the diagonal elements of the shear tensor ${\bf Q}$. In each mock, $\eta_t$ is known from the simulations and is not affected by any selection effects or measurement errors.
As shown in the top panel of Fig.~\ref{6dfmockqii}, the resultant $Q_{zz}$ has very large scatter about the true $Q_{zz}$, and the error bars are very large. By contrast, as show in the bottom panel of Fig.~\ref{6dfmockqii}, where $Q_{zz}$ is not an independent component (and calculated instead from $Q_{xx}$ and $Q_{yy}$ as in Eq.~\ref{tracQ}), the measured $Q_{zz}$ is consistent with the true values. 

\bsp	
\label{lastpage}
\end{document}